\documentclass[aps,pre,superscriptaddress
]{revtex4}

\usepackage{amsmath}
\usepackage{amssymb}
\usepackage{color}
\usepackage{float}
\usepackage{dcolumn}
\usepackage{bm}
\usepackage{newtxmath}
\usepackage{natbib}
\usepackage{hyperref}
\usepackage{graphicx}
\usepackage{CJKutf8}
\hypersetup
{
	colorlinks = true,
	urlcolor   = blue,
	citecolor  = blue,
	linkcolor  = blue,
}

\newcommand{\RomanNumeralCaps}[1]
\linenumbers

\begin{document}

\title{A unified theory for bubble dynamics}
\begin{CJK*}{UTF8}{gbsn}
	\author{A-Man Zhang (张阿漫)*}
	\email{zhangaman@hrbeu.edu.cn}
	\affiliation{College of Shipbuilding Engineering, Harbin Engineering University, Harbin, 150001, China}
	\affiliation{Nanhai Institute of Harbin Engineering University, Sanya, 572024, China}
	\author{Shi-Min Li  (李世民)}
	\affiliation{College of Shipbuilding Engineering, Harbin Engineering University, Harbin, 150001, China}
	\author{Pu Cui (崔璞)}
	\affiliation{College of Shipbuilding Engineering, Harbin Engineering University, Harbin, 150001, China}
	\affiliation{Nanhai Institute of Harbin Engineering University, Sanya, 572024, China}
	\author{Shuai Li (李帅)}
	\affiliation{College of Shipbuilding Engineering, Harbin Engineering University, Harbin, 150001, China}
	\affiliation{Nanhai Institute of Harbin Engineering University, Sanya, 572024, China}
	\author{Yun-Long Liu (刘云龙)}
	\affiliation{College of Shipbuilding Engineering, Harbin Engineering University, Harbin, 150001, China}
	\affiliation{Nanhai Institute of Harbin Engineering University, Sanya, 572024, China}
	\date{\today}
	
	\begin{abstract}
		In this work, we established a 
		novel theory for the dynamics of 
		oscillating bubbles such as cavitation bubbles, underwater explosion bubbles, and air bubbles. For the first time, we proposed bubble dynamics equations that can simultaneously take into consideration the effects of boundaries, bubble interaction, ambient flow field, gravity, bubble migration, fluid compressibility, viscosity, and surface tension while maintaining a unified and elegant mathematical form. The present theory unifies different classical bubble equations such as the Rayleigh-Plesset equation, the Gilmore equation, and the Keller-Miksis equation.
		Furthermore, we validated the theory with experimental data of bubbles with a variety in scales, sources, boundaries, and ambient conditions and showed the 
		advantages
		of our theory over the classical 
		theoretical models, followed by a discussion on the applicability of the present theory based on a comparison to simulation results with different numerical methods. Finally, as a demonstration of the potential of our theory, we modeled the complex multi-cycle bubble interaction with wide ranges of energy and phase differences and gained new physical insights into inter-bubble energy transfer and coupling of bubble-induced pressure waves. 
	\end{abstract}
	
	\maketitle
	
\end{CJK*}

\section{Introduction}
\label{sec:intro}

Bubbles are ubiquitous in nature and of great significance in numerous fields of science and engineering such as marine science,  ocean engineering, mechanical and material engineering, environmental and chemical engineering, and medicine and life science \citep{vshl00, sf08, vas12, Rod2015, Lohse2018D, Deike2022}. Bubble dynamics is a fundamental scientific problem attracting widespread interest. The pulsating bubbles that undergo volumetric oscillations due to pressure imbalance have many important applications. For example, some acoustic and laser cavitation bubbles \citep{bai2014,Movahed2016,Yuning2017,Bruning2019,Maeda2019,Hupfeld2020,Jingzhu2021,Kiyama2021}, which are typically millimeter or micrometer-sized oscillating bubbles, can facilitate ultrasonic cleaning \citep{Chahine2016,Landel2021}, sonoluminescence \citep{gaitan1992sonoluminescence,Akhatov1997,bhl02}, ultrasound therapy \citep{Lokhandwalla2001,oaijv2006,Lajoinie2014,Pahk2018,Mancia2020,Barney2020,Kikuchi2021,RN112021}, drug/gene delivery \citep{Brennen2015,lhgh18,sc19,Dollet2019,omata2022ultrasound}, inkjet printing \citep{Turkoz18,saade2021,Lohse2022}, and bubble propulsion \citep{liush17,Nourhani2020}. Underwater explosion/implosion bubbles and air-gun bubbles, which can be meter-sized oscillating bubbles, are central to underwater explosions/implosions \citep{Cole1948book,khwwk05,TurnerS2007,WangJ2018,LiS2019,wang2021theoretical,sun2022numerical} and geophysical explorations \citep{de2014bubble,Watson19,wehner2020acoustic}, respectively. Hydraulic cavitation bubbles \citep{JiB2015,Tairan2017,Wan2017,carling2017bubble,Sreedhar2017,Ludar2020,Aminoroayaie2021,Jeong2022}, entrained air bubbles/cavities \citep{Deane2008,c10,SeoJ2010,Roghair2011,Enriquez2012,Truscott2014,Smith2017}, heat-generated vapor bubbles \citep{Elbing2016,liu2017comparison,bhati2018semi}, and spark-generated bubbles \citep{okdbnf06,Dadvand2012,zccw15,Goh2017,kluesner2019practical,liang2021interaction}, which also oscillate and are usually sized from millimeters to meters, can be crucial to the performance of turbines, propellers, waterborne vehicles, engines, reactors, and spark sound sources. There are also oscillating bubbles with ultra small or large sizes, such as oscillating nanobubbles \citep{Menzl2016,ro21} or the bubbles produced in an underwater volcano eruption \citep{lhfww19}. 
Today, new studies on bubble dynamics keep emerging and new applications \citep{elze1999variational,Nguyen05,Amano11,spzh14,fogacca2016bubble,Wangy2018,Birjukovs2020,Murakami2021,chong2021improving,HaasT2021} of the oscillating bubbles are being discovered.

The dynamics of bubbles, of either natural or artificial sources, is complicated due to the existence of different scales, various boundaries, and multiple oscillation cycles that bring enormous challenges to theoretical, numerical, and experimental researches. Theoretical research is crucial to understanding the physics of bubbles under different conditions, which is the foundation for the utilization of bubbles and the circumvention of their hazardous effects. The theoretical research on bubble dynamics dates back to the early 20th century. \citet{r17} proposed a preliminary mathematical description of the collapse of a cavity in an infinite fluid field. Based on that, Plesset derived the classic Rayleigh-Plesset (RP) equation for bubble oscillation in an incompressible flow \citep{plesset49}. The loss in bubble energy due to acoustic radiation, such as bubble collapse-induced pressure waves \citep{Yasui1998,lt10,Supponen2016,Jaekyoon2018,Cui2020,LiangX2022,Trummler2020,Denner2023}, is overlooked due to the incompressible assumption and the RP equation may not be suitable when the damping in bubble radius, period, or energy is important. Taking into consideration the weak compressibility of the fluid outside the bubble, \citet{herring1941theory}, \citet{g52}, \citet{Trilling}, \citet{kk56}, \citet{Hickling}, \citet{Flynn1975}, \citet{km80}, and other researchers \citep{fa80} established weakly compressible bubble models. Among them, Keller's work in which the model was formulated using the wave equation and the incompressible Bernoulli equation enjoyed a most widespread application. Prosperetti and co-workers \citep{pp86, pp87} developed a model for bubble dynamics considering the compressibility of the far-field fluid based on the perturbation theory.
These models have played a major role in the theoretical research on bubble dynamics in a free field.
However, bubbles, regardless of type or scale, are seldom isolated and always in coupling with different conditions giving rise to complex bubble dynamics behaviors. In this paper, we proposed a theory describing bubble oscillation and migration in a compressible fluid. The theory provides a framework allowing for the effect of different types of boundaries, bubble interaction, background flow, gravity, migration, fluid compressibility, viscosity, and surface tension and presents a unification of important bubble models such as the Rayleigh-Plesset equation, the Gilmore equation, and the Keller-Miksis equation.

A major concern in the study of theoretical bubble models is the coupling effect of bubble oscillation and migration. \citet{h70} developed a bubble dynamics model considering the migration based on the incompressible assumption. \citet{Best1991} developed an incompressible model where a time-dependent dipole is inserted to allow for the migration of the spherical bubble. The migration-induced pressure variation was manifested by an additional term $v^2/4$ (surface-averaged pressure, SAP, and $v$ is the migration velocity). Similar methods were adopted by \citet{Brujan2005} and  \citet{SeoJ2010}. \citet{gh02} proposed a doubly asymptotic approximation (DAA) model specialized for a single underwater explosion bubble in a free field considering fluid compressibility and gravity-induced migration. To the best knowledge of the authors, extensions of the model to incorporate bubble migration for interacting bubbles or bubbles near boundaries in a compressible flow are still not available. 
The theory proposed in the current study can consider the effect of compressibility on not only bubble oscillation but also migration of various causation.

Another important issue for the theoretical bubble models is the interaction between multiple bubbles. Different bubble models have been applied to predict the behavior of a bubble in a multi-bubble system. \citet{Harkin2001} established a model for the spherical oscillation and translational motion of a pair of interacting gas bubbles in an incompressible liquid using Weiss sphere theorem 
and the method of SAP. \citet{Bremond-prl} extended the RP equation by considering the pressure modification induced by the surrounding bubbles and studied the weak interaction between two or more cavitation bubbles before the first bubble collapse. A similar scheme was also applied by \citet{gnadmo20} for cavitation bubbles and by \citet{Ziolkowski1982} and \citet{Li2014Liu} for airgun-bubble clusters in geophysical exploration. However, these previous works ignored the traveling time of pressure waves, which could be an oversimplification for the reproduction of the bubble-induced pressure field \citep{Grandjean2012,LiS2019,yunlong19,Beig2018}. To overcome this, we considered the time delay for the pressure wave propagation and our theory can reproduce some interesting phenomena that were not found with the previous models, such as the reflection of a pressure wave on bubble surfaces.

The effect of fluid boundaries on bubble behavior is another problem for the spherical bubble theories. To reproduce the bubble dynamics near a rigid wall, incompressible bubble models with the image method \citep{Best1991,Brujan2005} as well as the Gilmore model with the SAP method \citep{Chahine2009,Chahine2016} were used. The image method has also been applied to the free surface boundary \citep{zhangs2017}. For more accurate prediction, we incorporated in our model the effect of the bubble-induced velocity fields, pressure waves, and their reflections at the boundaries on the bubble behavior. This is based on consideration of fluid compressibility and traveling time of disturbances. It is worth noting that the negative pressure that is formed by the reflection of pressure pulses at the free surface can be captured by our theoretical model. 

This paper is structured as follows. The present theory for bubble dynamics, including the unified equations for bubble oscillation and migration, is derived in section \ref{sect:basic-theory}. The theory is extended to multiple-bubble interaction in section \ref{sect:theory-muti-bubble} and bubble dynamics near boundaries in section \ref{sect:theory-boundary}. A comparison with previous bubble models and experimental results with a variety in the scale and source of bubbles, boundaries, and other ambient conditions are presented in section \ref{sect:comparison}. A discussion on the applicability is given in section \ref{sect:discussion} followed by a demonstration of the capability of the present theory via solving complex interaction between two bubbles with phase and energy differences. Finally, this work is summarized and conclusions are drawn in section \ref{sect:conclusion}.

\section{Derivation of the unified theory for bubble dynamics}\label{sect:basic-theory}
\subsection {Basic equations of the theory}
Firstly, we derive the fundamental equation of the unified bubble dynamics theory from the basic laws of conservation. 
{
	The density, pressure, velocity, and deviatoric stress tensor of the fluid outside the bubble are denoted as $\rho$, $p$, $\boldsymbol{u}$, and $\boldsymbol{\tau}$, respectively. The conservation equations for mass and momentum are
	\begin{equation}
	\label{eq1}
	\frac{{\partial \rho }}{{\partial t}}  = - \nabla  \cdot \left( {\rho \boldsymbol{u}} \right)
	\end{equation}	
	
	\noindent and 
	
	\begin{equation}
	\label{eq2}
	\frac{\partial \boldsymbol{u}}{\partial t} + (\boldsymbol{u}\cdot\nabla)\boldsymbol{u} 
	= -\frac{1}{\rho}\nabla p + \frac{1}{\rho}\nabla \cdot \boldsymbol{\tau} +  \mathcal{F},
	\end{equation}
	where $\mathcal{F}$ is the body force. 
	Meanwhile, 
	the sound speed $C$ in the external fluid field 
	is related to the pressure $p$ and the fluid density $\rho$ as
	\begin{equation}
	\label{sound_speed}
	{C^2} = \frac{{{\rm{d}}p}}{{{\rm{d}}\rho }}.
	\end{equation}

	Substituting equations (\ref{eq1}) and (\ref{sound_speed}) into the derivative of equation (\ref{eq2}) with respect to time 
	yields

	\begin{equation}
	\label{eq4}
	\frac{1}{{{C^2}}}\frac{{{\partial ^2}\boldsymbol{u}}}{{\partial {t^2}}} + \frac{1}{{{C^2}}}\frac{\partial }{{\partial t}}\left[ {(\boldsymbol{u} \cdot \nabla )\boldsymbol{u}} \right]  =- \frac{1}{{{\rho ^2}}}\nabla  \cdot \left( {\rho {\boldsymbol{u}}} \right)\nabla \rho  + \frac{1}{\rho }\nabla \left[ {\nabla  \cdot \left( {\rho {\boldsymbol{u}}} \right)} \right] + \frac{1}{C^2}
	\frac{\partial ((\nabla \cdot \boldsymbol{\tau})/\rho +  \mathcal{F})}{\partial t}.
	\end{equation}
	\noindent 
	The sound speed $C$ is usually much greater than $|\boldsymbol{u}|$. Thus, we adopt the weakly compressible assumption and neglect the density gradient in the above equation. 
	The oscillating bubble driven by inertia is usually of large Reynolds numbers and short oscillating cycles. When only a few cycles are considered, the effect of viscosity can be neglected \cite{km80,pp86,wb10} here.  Thus, the first item of the last term in equation (\ref{eq4}) is dropped, and we will introduce the viscosity back to the system later from the dynamic boundary condition of the bubble surface.}
Assuming that the flow around the bubble is irrotational, then there exists the velocity potential $\varphi$ that satisfies $\boldsymbol{u} = \nabla \varphi$. Substituting it into equation (\ref{eq4}) and integrating from infinity to the current location while neglecting 
the variation of the body force and 
the terms containing $C^{-2}$ except the first one, we then have the wave equation
\begin{equation}
\label{eq-wave-equation}
\frac{1}{{{C^2}}}\frac{{{\partial ^2}\varphi }}{{\partial {t^2}}}   = {\nabla^2}\varphi.
\end{equation}

For an oscillating bubble, migration may happen due to the anisotropy of the surrounding flow such as the pressure gradient caused by nearby boundaries or the gravity field.
We tie the origin of a spherical coordinate system $o - r\theta \phi$ to the bubble center.
$\theta=0$ points to the direction of the bubble migration velocity relative to the ambient flow denoted by $\boldsymbol{v}$. 
$\boldsymbol{v} = v\boldsymbol{e} = {{\boldsymbol{v}_{\rm{m}}} - {{\boldsymbol{u}}_{\rm{a}}}}$, where ${\boldsymbol{e}}$ is a unit vector along the direction of $\boldsymbol{v}$, ${\boldsymbol{v}_{\rm{m}}}$ is the migration velocity, and ${{\boldsymbol{u}}_\mathrm{a} }$ is the ambient flow velocity. ${{\boldsymbol{u}}_\mathrm{a} } = {{\boldsymbol{u}}_\mathrm{B} } + {{\boldsymbol{u}}_\mathrm{E} }$, where ${{\boldsymbol{u}}_\mathrm{B} }$ is the velocity induced by boundaries or other bubbles and 
${{\boldsymbol{u}}_\mathrm{E} }$ represents extra velocities including the velocity of the incoming background flow ${{\boldsymbol{u}}_\infty}$.

In this work, the bubble with the radius of $R$ is assumed to oscillate and migrate while maintaining a spherical shape. We construct a solution of equation (\ref{eq-wave-equation}) with two singularities, i.e., a source and a dipole, the strength of which are denoted by $f$ and $q$, respectively. 
The velocity potential at an arbitrary location $\boldsymbol{r}$ and time $t$ 
induced by the moving singularities can be found in Appendix A. However, it is hard to write down explicit expressions when they are moving at varying speeds. 
If we consider the flow field within a small range around the bubble, despite the influences from boundaries and other bubbles, the direction of bubble migration can be deemed constant	and the velocity field axisymmetric to $\theta=0$. 
When $\boldsymbol{r}$ is close to the bubble and considering that the migration velocity $v$ is small compared with $C$, the variation of the location of the singularities can be neglected during the very short time when the influences are propagating from the singularities to an arbitrary location $\boldsymbol{r}$.
Thus, the velocity potential at $\boldsymbol{r} = (r,\theta)$ and $t$ is expressed as the summition of equations (\ref{eq-source-moving}) and (\ref{eq-dipole-moving}) after simplification, and reads
\begin{equation}
\label{eq-velocity-potential}
\varphi{(r,\theta,t)} = \frac{1}{r}f{\left(t-\frac{r}{C}\right)} + \frac{\mathcal{\cos\theta}}{r^2}q{\left(t-\frac{r}{C}\right)} + 
\frac{1}{C}\frac{\cos\theta}{r}q'{\left(t-\frac{r}{C}\right)}
\end{equation}
\noindent in which $q'$ denotes the derivative of $q$ with respect to its argument.



Note that, the wave equation governing the fluid field outside the bubble is not applicable to the fluid field inside the bubble and the propagation from the singularities to the bubble surface is a mathematical extension of the fluid field outside the bubble.
Then, we have the two velocity components at the bubble surface, induced by the two singularities, as
\begin{equation}
\label{eq-phi-r}
u_{r}{(R,\theta,t)} = 
\frac{\partial \varphi}{\partial r} = 
- \frac{f}{R^2}- 
\frac{1}{RC} f' 
-\frac{2\cos\theta}{R^3}q - 
\frac{2\cos\theta }{CR^2 }q'
\end{equation}
and
\begin{equation}
u_{\theta}{(R,\theta,t)} = 
\frac{1}{R}\frac{\partial \varphi}{\partial \theta} =
-\frac{\mathcal{\sin\theta}}{R^3}q
-\frac{1}{C}\frac{\sin\theta}{R^2}q',
\end{equation}
respectively, in which $f'$  denotes the derivative of $f$ with respect to its argument. The kinetic boundary condition on the bubble surface for the oscillation and the migration can be expressed as
\begin{equation}
\label{eq-kinetic-condition-osc}
\int_{\partial V} u_r \mathrm{d}S = 4\pi R^2\dot{R}
\end{equation}
and
\begin{equation}\label{eq-kinetic-condition-mig}
\frac{\mathrm{d}}{\mathrm{d}t}\int_V r\cos\theta \mathrm{d}V = \frac43\pi R^3v,
\end{equation}
respectively, where $V$ is the volume occupied by the bubble and $\partial V$ is its boundary, i.e., the bubble surface. $\dot{R}$ denotes the time derivative of $R$. 
Let's denote $F(t) = -f(t-R/C)$ and $Q(t) = q(t-R/C)$.
Substituting equation (\ref{eq-phi-r}) into equations (\ref{eq-kinetic-condition-osc}) and (\ref{eq-kinetic-condition-mig}), we have 
\begin{equation}
\label{eq-ODE-boundarycondition-osc}
F + \frac{R}{C-\dot{R}}\frac{\mathrm{d}F}{\mathrm{d}t} = R^2\dot{ R}
\end{equation}
and
\begin{equation}
\label{eq-ODE-boundarycondition-mig}
\frac{Q}{R}+\frac{1}{C-\dot{R}}Q'  = -\frac12 R^2v,
\end{equation}
respectively.
Employing the perturbation method on equation (\ref{eq-ODE-boundarycondition-mig}), we arrive at 
\begin{equation}\label{eq-s}
Q = -\frac{R^3}{2}v + 
\frac{R^3}{2C}\left(v\dot{R}+R\dot{v}\right)
\end{equation}
with the terms of the order $C^{-2}$ ignored.
Then, take the derivative of equation (\ref{eq-ODE-boundarycondition-osc}) with respect to $t$,  we have
\begin{equation}\label{eq-core-1}
\frac{\mathrm{d}F}{\mathrm{d}t} + \frac{\mathrm{d}}{\mathrm{d}t}\left(\frac{R}{C-\dot{R}}\frac{\mathrm{d}F}{\mathrm{d}t} \right) = 
2R\dot{R}^2 + R^2 \ddot{R}
\end{equation}
which may be reorganized as
\begin{equation}\label{eq-core-2}
\left( {\frac{{C - \dot R}}{R} + \frac{\mathrm{d}}{{\mathrm{d}t}}} \right)\left( \frac{{R}}{C - \dot R}\frac{\mathrm{d}F}{\mathrm{d}t} \right)= 2R{{\dot R}^2} + {R^2}\ddot R.
\end{equation}

Equation (\ref{eq-core-2}) describes the oscillation of a migrating bubble and is the base of the present unified 
theory for bubble dynamics. Its right-hand side equals $\ddot V/4\pi$, where $\ddot V$ is the bubble volume acceleration, meanwhile, the left-hand side is equivalent to the corresponding driving force. ${\mathrm{d}F}/{\mathrm{d}t}$ can be determined according to the physical problem considered, enabling equation (\ref{eq-core-2}) to model complicated bubble dynamics under different conditions. Therefore, the present theory can be extended while retaining a unified, concise, and elegant mathematical form to incorporate different conditions. The above forms the basis for the subsequent work of this paper. Additionally, if we consider the non-spherical motion of a simply-connected bubble, the velocity potential, and the bubble surface shape may be expanded by the bubble deformation modes. If the motion of a toroidal bubble (bubble ring) is considered, we may need to move the singularities off the rotating axis and introduce an extra vortex to incorporate the velocity circulation around the toroidal section.

\subsection {Bubble oscillation equation} 
In this section, we derive the bubble oscillation equation by introducing dynamic boundary conditions into equation (\ref{eq-core-2}). 
The deformation of fluid particles during the bubble oscillation would introduce additional viscous stress components. 
Viscosity was neglected in bulk liquid in the previous derivation. However, within the proximity of the bubble boundary, the viscosity effect can be restored as a correction term to  
the equilibrium condition for normal stresses on the bubble surface which involves the pressure of the internal gas on the inner bubble surface $P_\mathrm{g}$,  the fluid pressure on the outer bubble surface $P_{\rm{b}}$, surface tension, viscous stress, and other additional terms. Here we express the equilibrium condition as
\begin{equation}
\label{eq16}
{P_{\rm{g}}} = {P_{\rm{b}}} + \frac{{2\sigma }}{R} + \frac{{4\mu \dot R}}{R},
\end{equation}
where $\sigma$ is the surface tension and $\mu$ the liquid viscosity. 
$P_{\rm{g}}$ may be affected by many factors such as mass and heat transfer, wave effect, and chemical reactions, and can be determined according to the problem being considered. For simplicity, we assume that the oscillation process is adiabatic and that the internal gas pressure is uniform. Thus, $P_{\rm{g}}$ is composed of the non-condensable gas pressure and the saturated vapor pressure $P_{\rm{v}}$ as

\begin{equation}
\label{eq17}
{P_{\rm{g}}} = {P_0}{\left( {\frac{{{R_0}}}{R}} \right)^{3\gamma }} + {P_{\rm{v}}},
\end{equation}
where $P_0$ is the initial pressure of the non-condensable gas, $\gamma$ the polytropic exponent and $R_0$ the initial bubble radius.

To obtain ${\mathrm{d}F}/{\mathrm{d}t}$ in equation (\ref{eq-core-2}), we introduce the dynamic boundary condition expressed by the Bernoulli equation in the moving coordinate system

\begin{equation}
\label{eq-bernoulli}
\frac{ \partial \varphi}{ \partial t} - \boldsymbol{v} \cdot \boldsymbol{u}+ \frac{1}{2}{\left| \boldsymbol{u} \right|^2} + H = 0,
\end{equation}
\noindent where  $\boldsymbol{v} \cdot \boldsymbol{u}=\left( {v{\rm{cos}}\theta , - v{\rm{sin}}\theta } \right) \cdot \left( u_r,u_\theta \right)$ 
and ${\left| \boldsymbol{u} \right|^2}=  u_r^2 + u_\theta^2$. 
$H=\int_{{P_{\rm{a}}}}^{{P_{\rm{b}}}} {\rho^{-1}} {} \mathrm{d} p$ is the enthalpy difference at the bubble surface. Here, $P_\mathrm{a}$ represents the ambient pressure at the bubble center. $P_\mathrm{a}=P_\mathrm{B}+P_\mathrm{E}$ where  $P_\mathrm{B}$  is the pressure induced by boundaries or other bubbles and $P_\mathrm{E}$ represents extra pressures including the hydrostatic pressure $P_\infty$ and acoustic pressures $P_\mathrm{A}$. 
With the motion of the fluid being an adiabatic process, $H$ resolves into 
\begin{equation}
\label{H-epsilon-0}
H =  C^2\varpi -\frac12 C^2 \varpi^2  + O(\varpi^3),
\end{equation}
where $\varpi = (P_\mathrm{b}-P_\mathrm{a})/(\rho C^2)$.  


Deriving equation (\ref{eq-velocity-potential}) with respect to time and substituting it into equation (\ref{eq-bernoulli}). Then, combining the relationship between $F$ and $f$ can produce $\mathrm{d}F/\mathrm{d}t$. Note that the terms containing $\theta$ in equation (\ref{eq-bernoulli}) will lead to non-spherical bubble deformation. To comply with the spherical assumption while taking into consideration their averaged effects, we integrate equation (\ref{eq-bernoulli}) over the bubble surface to eliminate $\theta$. Then, $\mathrm{d}F/\mathrm{d}t$ could be obtained
\begin{equation}
\label{cm20}
\frac{\mathrm{d}F}{\mathrm{d}t} 
=  R \left( {1 - \frac{{\dot R}}{C}} \right) \left( {\frac{1}{2}{{\dot R}^2} + \frac{v^2}{4} + H} \right).
\end{equation}

Finally, substitute the above into equation (\ref{eq-core-2}) and we have the bubble oscillation equation 
\begin{equation}
\label{eq33}
{\color{black}\left( {\frac{{C - \dot R}}{R} + \frac{\mathrm{d}}{{\mathrm{d}t}}} \right)\left[ {\frac{{{R^2}}}{C}\left( {\frac{1}{2}{{\dot R}^2} + \frac{1}{4}{v^2} + H} \right)} \right]  = 2R{{\dot R}^2} + {R^2}\ddot R.}
\end{equation}

Equation (\ref{eq33}) is the core of the present theory 
that can be extended to various scenarios such as multiple-bubble interaction and bubble dynamics near boundaries.
The above equation evolves into a unified and elegant mathematical form similar to equation (\ref{eq-core-2}) as
\begin{equation}
\label{coref}
\frac{1}{C}\frac{\mathrm{d}\tilde{F}}{\mathrm{d}t}+
\left(1-\frac{\dot{R}}{C}\right)\frac{\tilde{F}}{R} = 
2R{\dot R^2} + {R^2}\ddot R , 
\end{equation}
where $ \tilde{F} = R^2 (\dot{R}^2/2+v^2/4+H)$ is to be determined according to specific conditions of the problem under consideration.
$\dot{R}^2/2$, $v^2/4$, and $H$ indicate equivalent forces induced by bubble oscillation, migration, and the ambient flow field, respectively. 

The oscillation in a gravity field is described by equation (\ref{eq33}) with the ambient flow ignored  ($ \boldsymbol{u}_\mathrm{a} =0 $).
In addition, neglecting both the ambient flow and bubble migration ( i.e. $\boldsymbol{v} = 0 $), equation (\ref{eq33}) is simplified as 
\begin{equation}\label{eq-core-n}
\left( {\frac{{C - \dot R}}{R} + \frac{{\rm{d}}}{{{\rm{d}}t}}} \right)\left[ {\frac{{{R^2}}}{C}\left( {\frac{1}{2}{{\dot R}^2} + H} \right)} \right] = 2R{\dot R^2} + {R^2}\ddot R,
\end{equation}
which describes bubble oscillation in a free field. 
Equation (\ref{eq33}) can be expanded as the following form

\begin{equation}
\left(1+\frac{ \dot{R}}{C} \right) H+
\frac{R}{ C} \dot{H}
+\frac14 \left(1+\frac{\dot{R}}{C}\right)v^2 + \frac{{R}}{2C}v\dot{v} \\
=
\frac{3}{2} \left(1-\frac{ \dot{R}}{3C} \right)\dot{R}^2 +
\left(1-\frac{ \dot{R}}{C} \right) R\ddot{R}.
\label{Equation:Keller1}
\end{equation}
It can be  simplified to the Keller-Miksis equation when $v=0$, $P_{\mathrm{a}} = P_{\infty}$, and $H$ is calculated with equation (\ref{H-epsilon-0}) retaining only the first term. 
On this basis,
if the second term on the left-hand side is substituted with
$(1-\dot{R}/C)R\dot{H}/C$, equation (\ref{Equation:Keller1}) becomes the Gilmore equation.




For bubble dynamics in an incompressible flow, i.e., $ C\rightarrow \infty $, equation (\ref{eq-core-2}) can be reduced to
\begin{equation}\label{eq-core-1fu1}
\frac{\mathrm{d}F}{\mathrm{d}t} = 
2R\dot{R}^2 + R^2 \ddot{R},
\end{equation}
and equation (\ref{eq33}) can be simplified as
\begin{equation}
\label{eq33simple}
\frac{1}{4}v^2 + \frac{{{P_{\rm{b}}} - {P_\mathrm{a}}}}{\rho } = \frac{3}{2}{\dot R^2} + R\ddot R,
\end{equation}
which further reduces to the Rayleigh-Plesset equation when the first term is neglected and $P_{\mathrm{a}} = P_{\infty}$. 
{Therefore, our theory unifies the Rayleigh-Plesset equation, the Gilmore equation, and the Keller-Miksis equation.
}

\subsection{Bubble migration equation}
\label{Sect:migration}

Since the problem is migration-related, equation (\ref{eq33}) involves two unknowns, $R$ and $v$. Therefore, an additional equation is required to make the problem solvable. 
The momentum equation of the bubble migration in the gravity field can be written as
\begin{equation}
\label{eq-migration-momentum}
m \dot {\boldsymbol{v}}_\mathrm{m} =  m {\boldsymbol{g}} - \int_S {{P_b}{\boldsymbol{n}}\mathrm{d}s}  - \frac{1}{2}\rho {A_\mathrm{p}}{C_\mathrm{d}}\mathbb{S} \boldsymbol{v} + \boldsymbol{X},
\end{equation}
where $m$ is the total mass of the gas inside the bubble, $\boldsymbol{g}$ is the gravity acceleration, $ C_{\rm{d}} $ is the drag coefficient, $A_{\rm{p}}$ is the projected area of the bubble in the migration direction, and $ \mathbb{S}(\cdot) = (\cdot)|\cdot| $ is a signed square operator. The second term on the right-hand side denotes the inviscid part of the hydrodynamic force exerted  on the bubble by the surrounding fluid and the third term approximates the viscous part. $\boldsymbol{X}$ represents extra forces such as lift force and is not considered in this paper.

Substituting $H=\left({P_{\rm{b}}-P_{\rm{a}}}\right)/{\rho}$ into equation (\ref{eq-bernoulli}) will give us $P_\mathrm{b}$. Then, the inviscid resultant force on the bubble can be 
expressed as
\begin{equation}
\label{eq-inviscid-force-migration}
- \int_S {{P_b}{\boldsymbol{n}}\mathrm{d}s}  = -\rho  \frac{\mathrm{d}\left( C_{\rm{a}}V \boldsymbol{v} \right)}{\mathrm{d}t} - 
V\nabla {P_\mathrm{a}},
\end{equation}
where $C_\mathrm{a}$ is the added mass coefficient and $V = 4/3\pi R^3$ is the bubble volume and $\boldsymbol{n}$ is the external normal vector.

Since the density of the gas inside the bubble is usually much smaller than that of the liquid outside the bubble, we may ignore the inertial force and the gravity force of the gas. Hence, equation (\ref{eq-migration-momentum}) can be transformed into
\begin{equation}
\label{eq-migration}
{C_\mathrm{a}} R \dot {\boldsymbol{v}} + \left( {3{C_\mathrm{a}}\dot R + {{\dot C}_\mathrm{a}}R} \right) \boldsymbol{v} 
+ \frac{R}{\rho }\nabla {P_\mathrm{a}} + \frac{3}{8}{C_\mathrm{d}}\mathbb{S} \left(\boldsymbol{v}\right) = 0.
\end{equation}

For a bubble oscillating and migrating in a still and unbounded flow,  we have $\boldsymbol{v}  = \boldsymbol{v}_\mathrm{m} $ and $\nabla {P_{\rm{a}}} = \rho \boldsymbol{g} = (0,0,-\rho g)$. Then equation (\ref{eq-migration}) reduces to
\begin{equation}
\label{eq36}
C_\mathrm{a}R\dot{v}_\mathrm{m} + (3C_\mathrm{a}\dot{R} +  \dot{C}_\mathrm{a}R) v_\mathrm{m}
- 
gR + \frac38{C_{\rm{d}}}\mathbb{S} (v_\mathrm{m}) =0.
\end{equation}

The bubble oscillation and migration can be obtained by solving equations (\ref{eq33}) and (\ref{eq-migration}) simultaneously, or by solving (\ref{eq33}) and (\ref{eq36}) if the ambient flow is not considered.	

\subsection{Bubble-induced velocity and pressure fields}
We can calculate the velocity and pressure fields induced by the bubble once the bubble motion is solved. 
The Bernoulli equation at an arbitrary location $\boldsymbol{r}$ reads
\begin{equation}
\label{eq22}
\frac{{\partial \varphi }}{{\partial t}}  + \frac{1}{2}
(
|\boldsymbol{u}|^2 
- |\boldsymbol{u}_\mathrm{a}|^2
)
+ \frac{p - P_\mathrm{a}}{\rho }= 0,
\end{equation}

\noindent where the third term is $H$ calculated by equation (\ref{H-epsilon-0}) retaining only the first term.
Combining equations (\ref{eq-velocity-potential}) and (\ref{cm20}), 
one may obtain the relation between the physical quantities at $\boldsymbol{r}$ and at the bubble surface as
\begin{equation}
\label{eq24}
\boldsymbol{u}_{\left( {\boldsymbol{r},t} \right)} = \frac{\boldsymbol{r}R}{|\boldsymbol{r}|^3}\left. {\left( {R\dot R - \frac{|\boldsymbol{r}|-R}{C}\frac{{\partial \varphi }}{{\partial t}}} \right)} \right|_{\left( {R,t - \frac{{|\boldsymbol{r}| - R}}{C}} \right)}
\end{equation}
\noindent and
\begin{equation}
\label{eq23}
{\left. {\frac{{\partial \varphi }}{{\partial t}}} \right|_{\left( {\boldsymbol{r},t} \right)}} = {\left. { - \frac{R}{|\boldsymbol{r}|}\left( {H + \frac{1}{2}{{\dot R}^2}} +\frac{1}{4}v^2\right)} \right|_{\left( {R,t - \frac{{|\boldsymbol{r}| - R}}{C}} \right)}},
\end{equation}
where the higher order terms of $|\boldsymbol{r}|^{-1}$ and the effect of migration are ignored. Note that $R$ on the right-hand side of the above two equations including those in the subscripts should also be evaluated at  
$t - (|\boldsymbol{r}|-R)/C$. Here, $|\boldsymbol{r}|-R$ denotes the minimum distance from the bubble surface to $\boldsymbol{r}$.
Substituting the above into equation (\ref{eq22}) and ignoring the ambient velocity $\boldsymbol{u}_\mathrm{a}$, one may obtain the formula for the pressure at $\boldsymbol{r}$ as

	\begin{equation}
	\label{eq25}
	p(\boldsymbol{r},t) =  P_\mathrm{a} +  \rho \left.
	\left\{
	\frac{R}{|\boldsymbol{r}|}\left(H+\frac12\dot{R}^2+\frac{1}{4}v^2\right) -
	\frac12\frac{R^2}{|\boldsymbol{r}|^4}
	\left[R\dot{R}+\frac{|\boldsymbol{r}|-R}{C}\left(H+\frac12 \dot{R}^2+\frac{1}{4}v^2\right)
	\right]^2
	\right\}
	\right|_{\left(R,t-\frac{|\boldsymbol{r}|-R}{C}\right)}
	. 
	\end{equation}

Note that, the pressure disturbance in the fluid field with a distance $r$ from the bubble center at time $t$ is induced by the bubble at an earlier time, $t-(|\boldsymbol{r}|-R)/C$, i.e., the bubble-induced disturbance is delayed in time due to fluid compressibility. Ignoring the effect of bubble migration, in an incompressible flow where $C$ approaches infinity, we have
\begin{equation}
\boldsymbol{u}{(\boldsymbol{r},t)} = \frac{\boldsymbol{r}}{|\boldsymbol{r}|^3}R^2\dot{R}
\end{equation}
and
\begin{equation}
\left.\frac{\partial \varphi}{\partial t}\right|_{(\boldsymbol{r},t )} = 
- \frac{1}{|\boldsymbol{r}|} \frac{\mathrm{d}R^2 \dot{R}}{\mathrm{d}t}.
\end{equation}
Similarly, substituting them into equation (\ref{eq22}), we have the degraded counterpart of equation (\ref{eq25}) in an incompressible fluid expressed as
\begin{equation}
\label{eq26}
p{\left( {\boldsymbol{r},t} \right)} =
- \rho \frac{{{R^4}{{\dot R}^2}}}{{2{|\boldsymbol{r}|^4}}}
+ \rho \frac{{2R{{\dot R}^2} + {R^2}\ddot R}}{|\boldsymbol{r}|} 
+P_\mathrm{a}
.
\end{equation} 

\section{Modeling multiple-bubble interaction}\label{sect:theory-muti-bubble}

Bubbles usually do not appear in isolation either in nature or in engineering applications and the interaction between multiple bubbles may result in a significant alteration of the bubble behavior. Here, we derive the equations for the oscillation and migration of multiple bubbles based on the theory in Section \ref{sect:basic-theory}. 
For an arbitrary point $\boldsymbol{r}$ in the fluid field, the velocity potential and the velocity induced by the migration of bubble $N$ are of the orders of  $|\boldsymbol{o}_N - \boldsymbol{r}|^{-2}$ and $|\boldsymbol{o}_N - \boldsymbol{r}|^{-4}$, respectively, where $ \boldsymbol{o}_N $ is the center of bubble $ N $. Thus, only the effects of the oscillation of bubble $N$ need to be considered.  Hence, by taking the derivatives of equation (\ref{eq-velocity-potential}) with respect to $\boldsymbol{r}$ and $t$ with the solution of $f$ substituted, the current velocity and the time derivative of the potential induced by bubble $N$ at $\boldsymbol{r}$ can be expressed as
\begin{small}
	\begin{equation}
	\boldsymbol{u}_{N}{(\boldsymbol{r},t)} = 
	-\frac{\boldsymbol{o}_N - \boldsymbol{r}}{|\boldsymbol{o}_N - \boldsymbol{r}|^3}\left.
	\left[\frac{R_N}{C}\left(|\boldsymbol{o}_N - \boldsymbol{r}|-R_N\right)\left(H+\frac12\dot{R}_N^2+\frac14 v_N^2\right)	
	+R_N^2\dot{R}_N
	\right]
	\right|_{\left({R_{N}},t_N\right)}
	\end{equation}
\end{small}

\noindent and

\begin{equation}
\dot{\varphi}_{N}{(\boldsymbol{r},t)} = 
-\left. \frac{R_N}{|\boldsymbol{o}_N - \boldsymbol{r}|} \left( H + \frac12 \dot{R}_N^2+\frac14 v_N^2\right) \right|_{\left({R_{N}},t_N\right)},
\end{equation}

\noindent respectively, where $ t_N = t - (|{\boldsymbol{r}_{N}}| - R_N)/C $ is the initiation time of a disturbance induced by bubble $ N $ that later arrives at $ \boldsymbol{r} $ at time $ t $. In the equations, the subscript $M$ or $N$ indicates the quantities of the corresponding bubble. When considering the dynamics of bubble $M$, we interpret the effect of other bubbles on its ambient fluid field as disturbances in the velocity and the pressure fields. Thus, we may obtain the flow velocity $\boldsymbol{u}_\mathrm{B}$ and the pressure $P_\mathrm{B}$ of bubble $ M $  induced by other bubbles, considering the influences from all the other bubbles, as 

\begin{equation}
\label{eq39}
{\boldsymbol{u}}_{\mathrm{B}}{ ( {{{\boldsymbol{o}}_M},t} )}=\sum_{ N = 1,L\atop
	N \ne M} \boldsymbol{u}_N(\boldsymbol{o}_{M},t)
\end{equation}

\noindent and 

\begin{equation}
\label{eq37}
P_{\mathrm{B}}{ (\boldsymbol{o}_M,t)} = 
- \rho \sum_{ N = 1,L\atop
	N \ne M} \dot{\varphi}_N (\boldsymbol{o}_{M},t) - 
\frac12 \rho |\boldsymbol{u}_\mathrm{B}(\boldsymbol{o}_{M},t)|^2,
\end{equation}

\noindent where $L$ is the total number of the interacting bubbles and $\boldsymbol{o}_M $ is the center of bubble $ M $. Substituting the above two equations 
into equations (\ref{eq33})  and (\ref{eq-migration}) {for bubble $M$}, and setting $M$ as 1, 2, …, and $L$, respectively, one may obtain a set of 2$L$ equations that describe the oscillation and migration of the $L$ bubbles considering their mutual interaction. 
The pressure in the fluid field contains contributions from all the bubbles, therefore, the pressure at $\boldsymbol{r}$ can be expressed as

\begin{equation}
p{(\boldsymbol{r},t)}  = 	- 
\rho \sum_{ N = 1,L} \dot{\varphi}{N(\boldsymbol{r},t)} -
\frac12 \rho \left|\sum_{ N = 1,L} \boldsymbol{u}_{N}{(\boldsymbol{r},t)}\right|^2 
+P_\mathrm{E}
.
\end{equation}

When $C$ approaches infinity, i.e., the external fluid is incompressible, we can obtain the reduced bubble dynamics equations as

\begin{equation}
\label{eq43}
\frac{1}{4}{\left| {{{\boldsymbol{v}}_M} + \sum_{ N = 1,Q\atop
			N \ne M} {{{\boldsymbol{r}}_{MN}}\frac{{R_N^2{{\dot R}_N}}}{{{{\left| {{{\boldsymbol{r}}_{MN}}} \right|}^3}}}} } \right|^2}  
- 
\sum_{ N = 1,Q\atop
	N \ne M} {\left( {\frac{{2{R_N}\dot R_N^2 + R_N^2{{\ddot R}_N}}}{{{{\left |{\boldsymbol{r}}_{MN} \right | }}}}} \right)} 
+\frac{{{P_{\rm{b}}} - {P_\mathrm{E} }}}{\rho } = 
\frac{3}{2}{\dot R_M^2} + R_M\ddot R_M,
\end{equation}
\noindent and
\begin{equation}\label{eq44}
\sum_{ N = 1,Q\atop
	N \ne M} {{{\boldsymbol{r}}_{MN}}\frac{{(2+5C_{\rm{a}})R_N^2\dot R_N^2 + (1+C_{\rm{a}})R_N^3{{\ddot R}_N}}+\dot{C}_{\rm{a}} R_N^3 \dot R_N}{{{{\left| {{{\boldsymbol{r}}_{MN}}} \right|}^3}}}}  -
\frac38{C_{\rm{d}}}\mathbb{S} 
\left( {{{\boldsymbol{v}}_{M}} + \sum_{ N = 1,Q\atop
		N \ne M} {{{\boldsymbol{r}}_{MN}}\frac{{R_N^2{{\dot R}_N}}}{{{{\left| {{{\boldsymbol{r}}_{MN}}} \right|}^3}}}} } \right) 
+ R{\boldsymbol{g}}= C_{\rm{a}}R {\dot{\boldsymbol{ v}}_M} + (3C_{\rm{a}}\dot R + \dot{C_{\rm{a}}}R){{\boldsymbol{v}}_M},
\end{equation}

\noindent {where $ \boldsymbol{r}_{MN} = \boldsymbol{o}_N - \boldsymbol{o}_M $.}
The above equations with $M$ = 1, 2, …, and $L$ describe the dynamics of the multiple interacting bubbles in an incompressible flow. Particularly, for the dynamics of two interacting bubbles, i.e., $L=2$ and ($M$, $N$) = (1, 2) or (2, 1), the above equations reduce to

\begin{equation}
\label{eq43simple}
\frac{1}{4}{\left| {{\boldsymbol{v}_M} + {\boldsymbol{r}_{MN}}\frac{{R_N^2{{\dot R}_N}}}{{{{\left| {{{\boldsymbol{r}}_{MN}}} \right|}^3}}}} \right|^2}  - \frac{{2{R_N}\dot R_N^2 + R_N^2{{\ddot R}_N}}}{{\left| {{\boldsymbol{r}_{MN}}} \right|}} + \frac{{{P_\mathrm{b}} - {P_\mathrm{E} }}}{\rho }  = \frac{3}{2}{\dot R^2}_M + {R_M}{\ddot R_M},
\end{equation}

\noindent and

\begin{equation}
\label{eq44simple}
{{\boldsymbol{r}}_{MN}}\frac{{(2+5C_{\rm{a}})R_N^2\dot R_N^2 + (1+C_{\rm{a}})R_N^3{{\ddot R}_N}}+\dot{C}_{\rm{a}} R_N^3 \dot R_N }{{{{\left| {{{\boldsymbol{r}}_{MN}}} \right|}^3}}} -
\frac38{C_{\rm{d}}}\mathbb{S} \left( {{{\boldsymbol{v}}_{\rm{M}}} + {{\boldsymbol{r}}_{MN}}\frac{{R_N^2{{\dot R}_N}}}{{{{\left| {{{\boldsymbol{r}}_{MN}}} \right|}^3}}}} \right) + R{\bf{g}} 
={C_{\rm{a}}R {\dot{\boldsymbol{ v}}_M} + (3C_{\rm{a}}\dot R + \dot{C_{\rm{a}}}R){{\boldsymbol{v}}_M}}.
\end{equation}

\section{Modeling bubble dynamics near boundaries}\label{sect:theory-boundary}
Besides gravity and bubble interaction, nearby boundaries such as a rigid/solid wall or a free surface may also impose considerable influence on the bubble dynamics. 
In this paper, we incorporate the boundary effects by adopting the image theory in which the boundary condition can be satisfied by properly introducing image bubbles on the opposite side of the boundary. Then, the dynamics of a bubble is subjected to the influences of other real bubbles and all the image bubbles.
Thus, we derive the equations of bubble oscillation and migration subject to the boundary effect by transferring the problem to a special multi-bubble interaction problem.
Let us describe the boundary plane with ${{\boldsymbol{e}}_{\rm{b}}} \cdot {\boldsymbol{r}} + b = 0$ where ${\boldsymbol{e}}_{\rm{b}}$ is the outward normal vector of the boundary and $b$ is a constant. Assume that the boundary can reflect the influence of a bubble with a reflection coefficient denoted by $\alpha$ depending on the property of the boundary. Particularly, for a rigid boundary, $\alpha = 1$ and for an ideal free surface, $\alpha = -1$. For bubble $N_i$, which is the image of bubble $N$, it is required that 
${{\boldsymbol{o}}_{N_i}} = {{\boldsymbol{o}}_N} - \left( 2{{{\boldsymbol{o}}_N} \cdot {{\boldsymbol{e}}_{\rm{b}}}  + b} \right){{\boldsymbol{e}}_{\rm{b}}}$, ${R_{N_i}} = {R_N},{\dot R_{N_i}} = \alpha {\dot R_N}$, ${\ddot R_{N_i}} = \alpha {\ddot R_N}$, and ${{\boldsymbol{v}}_{N_i}} = \alpha \left[ {{{\boldsymbol{v}}_N} - 2\left( {{{\boldsymbol{v}}_N} \cdot {{\boldsymbol{e}}_{\rm{b}}}} \right){{\boldsymbol{e}}_{\rm{b}}}} \right]$, respectively. Therefore, similar to equations (\ref{eq39}) and (\ref{eq37}), the flow velocity and the pressure at $\boldsymbol{o}_M$ induced by other bubbles and all the images can be written as

\begin{equation}
\label{eq47}
\boldsymbol{u}_{\mathrm{B}}{\left( {{{\boldsymbol{o}}_M},t} \right)}=
\sum_{\genfrac{}{}{0pt}{2}{N=1,L}{N\ne M}} \boldsymbol{u}_{N}{(\boldsymbol{o}_M,t)} + \sum_{N=1,L} \boldsymbol{u}_{{N_i}}{(\boldsymbol{o}_M,t)}
\end{equation}

\noindent and

\begin{small}
	\begin{equation}
	\label{eq-pa-boundary}
	P_{\mathrm{B}}{(\boldsymbol{o}_M,t)} =
	- 
	\rho \sum_{\genfrac{}{}{0pt}{2}{N=1,L}{N\ne M}} \dot{\varphi}_{N}{(\boldsymbol{o}_M,t)} 
	-\rho \sum_{N=1,L} \dot{\varphi}_{{N_i}}{(\boldsymbol{o}_M,t)}
	-\frac12 \rho |\boldsymbol{u}_{\mathrm{B}}{(\boldsymbol{o}_M,t)}|^2.
	\end{equation}
\end{small}

Substitute the above two equations
in to equations (\ref{eq33}) and (\ref{eq-migration}) for bubble $M$, we then have the equations for the dynamics of one or more bubbles considering boundary effect in a compressible flow. The above derivation can be further extended to various scenarios such as those with flexible, multiple, or hybrid boundaries. 
The pressure in the fluid field contains the contributions from all the bubbles and their images. Then the pressure at $\boldsymbol{r}$ can be written as

\begin{equation}
\label{eq48}
p{(\boldsymbol{r},t)} = 
-\rho \sum_{N=1,Q} \left[ \dot{\varphi}_{N}{(\boldsymbol{r},t)}+\dot{\varphi}_{{N_i}}{(\boldsymbol{r},t)}\right]
-
\frac12 \rho
\left|\sum_{N=1,Q} [\boldsymbol{u}_{N}{(\boldsymbol{r},t)} + \boldsymbol{u}_{{N_i}}{(\boldsymbol{r},t)}]\right|^2
+P_\mathrm{E}
.
\end{equation}
Note that, when the reflection coefficient $\alpha$ of the boundary is negative, the pressure waves induced by the image bubbles become rarefaction waves
that may fall within the cavitation limit of the liquid.
Therefore, it may be necessary to apply a modification, e.g., the cutoff cavitation model, to the pressure to incorporate the cavitation effect.
When $C$ approaches infinity, i.e., the fluid outside the bubble becomes incompressible, the oscillation and migration equations can be rewritten as

\begin{multline}
\label{eq49}
\frac14{\left| {{{\boldsymbol{v}}_M} + \sum_{ N = 1,Q\atop
			N \ne M} {{{\boldsymbol{r}}_{MN}}\frac{{R_N^2{{\dot R}_N}}}{{{{\left| {{{\boldsymbol{r}}_{MN}}} \right|}^3}}} + \sum_{N = 1,Q} {{{\boldsymbol{r}}_{MN_i}}\frac{{R_{N_i}^2{{\dot R}_{N_i}}}}{{{{\left| {{{\boldsymbol{r}}_{MN_i}}} \right|}^3}}}} } } \right|^2} \\ - 
\sum_{ N = 1,Q\atop
	N \ne M} {\left( {\frac{{2{R_N}\dot R_N^2 + R_N^2{{\ddot R}_N}}}{{\left| {{{\boldsymbol{r}}_{MN}}} \right|}}} \right)}- \sum_{N = 1,Q} {\left( {\frac{{2{R_{N_i}}\dot R_{N_i}^2 + R_{N_i}^2{{\ddot R}_{N_i}}}}{{\left| {{{\boldsymbol{r}}_{MN_i}}} \right|}}} \right)}   
+ \frac{{{P_{\rm{b}}} - {P_\mathrm{E} }}}{\rho }= \frac{3}{2}{{\dot R}^2}_M + {R_M}{{\ddot R}_M},
\end{multline}

\noindent and 
\begin{multline}
\label{eq50}
\sum_{ N = 1,Q\atop
	N \ne M} {{{\boldsymbol{r}}_{MN}}\frac{{(2+5C_{\rm{a}})R_N^2\dot R_N^2 + (1+C_{\rm{a}})R_N^3{{\ddot R}_N}}+\dot{C}_{\rm{a}} R_N^3 \dot R_N}{{{{\left| {{{\boldsymbol{r}}_{MN}}} \right|}^3}}}-\sum_{N = 1,Q} {{{\boldsymbol{r}}_{MN_i}}\frac{{(2+5C_{\rm{a}})R_{N_i}^2\dot R_{N_i}^2 + (1+C_{\rm{a}})R_{N_i}^3{{\ddot R}_{N_i}}}+\dot{C}_{\rm{a}} R_{N_i}^3 \dot R_{N_i}}{{{{\left| {{{\boldsymbol{r}}_{MN_i}}} \right|}^3}}}} }\\-\frac38{C_{\rm{d}}}\mathbb{S} \left( {{{\boldsymbol{v}}_M}+\sum_{ N = 1,Q\atop
		N \ne M} {{{\boldsymbol{r}}_{MN}}\frac{{R_N^2{{\dot R}_N}}}{{{{\left| {{{\boldsymbol{r}}_{MN}}} \right|}^3}}}+\sum_{N = 1,Q} {{{\boldsymbol{r}}_{MN_i}}\frac{{R_{N_i}^2{{\dot R}_{N_i}}}}{{{{\left| {{{\boldsymbol{r}}_{MN_i}}} \right|}^3}}}} } } \right) +
R{\boldsymbol{g}}
=C_{\rm{a}}R {\dot{\boldsymbol{ v}}_M} + (3C_{\rm{a}}\dot R + \dot{C_{\rm{a}}}R){{\boldsymbol{v}}_M},
\end{multline}
and one may then obtain the bubble dynamics considering the boundary effect and bubble interaction in an incompressible flow.  Particularly, for the dynamics of a single bubble interacting with a boundary, the above equations further reduce to 

\begin{equation}
\label{eq49simple}
\frac{1}{4}{\left| {{\boldsymbol{v_M}} +  {{\boldsymbol{e}}_\mathrm{b}}\frac{{R_i^2{{\dot R}_i}}}{{{4d_\mathrm{b}^2}}}} \right|^2}  -  \frac{{2{R_i}\dot R_i^2 + R_i^2{{\ddot R}_i}}}{2d_\mathrm{b} }+ \frac{{P_\mathrm{b}} - {P_\mathrm{E}}}{\rho } = \frac{3}{2}{\dot R^2} + R\ddot R
\end{equation}
and
\begin{equation}
\label{eq50simple}
{{\boldsymbol{e}}_\mathrm{b}}\frac{{(2+5C_{\rm{a}})R_i^2\dot R_i^2 + (1+C_{\rm{a}})R_i^3{{\ddot R}_i}}+\dot{C}_{\rm{a}} R_i^3 \dot R_i}{{{4d_\mathrm{b}^2}}}  - \frac38{C_{\rm{d}}}\mathbb{S} \left( {{\boldsymbol{v_M}} +  {{\boldsymbol{e}}_b}\frac{{R_i^2{{\dot R}_i}}}{{{4d_\mathrm{b}^2}}}} \right) + R{\boldsymbol{g}}  
=C_{\rm{a}}R {\dot{\boldsymbol{ v}}_M} + (3C_{\rm{a}}\dot R + \dot{C_{\rm{a}}}R){{\boldsymbol{v}}_M},
\end{equation}
where $ d_\mathrm{b} $ is the stand-off distance from the bubble center to the boundary.

\section{Comparison between theoretical and experimental results}\label{sect:comparison}
In this part, we validate our unified 
theory with experiments of acoustic bubbles, electric spark bubbles, underwater explosion bubbles, laser-induced cavitation bubbles, and air-gun bubbles that are varied in bubble scales, sources, boundaries, and other ambient conditions, as well as the results of classical bubble theories to demonstrate the advantages of the present theory.

\subsection{Bubble dynamics in a free field}
\begin{figure*}
	\centering\includegraphics[width=0.7\linewidth]{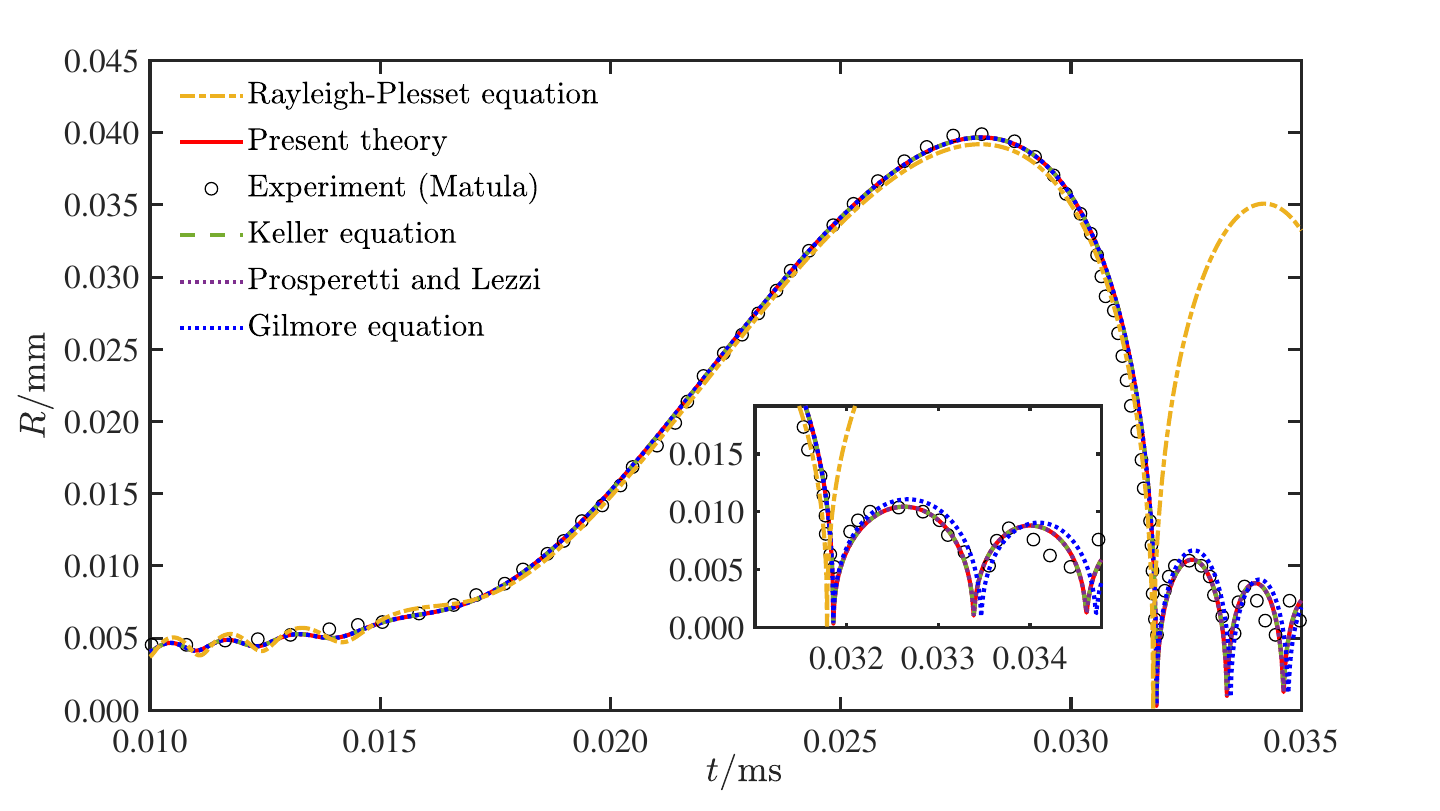}
	\caption{{A comparison between the theoretical results and the experimental data \citep{m99} of the radius of an acoustic bubble in a free field.}  }
	\label{figure1}
\end{figure*}

Firstly, we verify the correctness and validity of our unified theory for bubble dynamics using the experimental data of single sonoluminescence cavitation bubbles from \citet{m99} and results of the classical bubble theories from  \citet{plesset49}, \citet{g52}, \citet{km80}, and \citet{pp86} concerning the dynamics of a bubble in a free field. In the experiment, the bubble was initially in an equilibrium state with a radius of 4.7 ${\rm\mu}$m before being stimulated at $t = 0$ by a high-frequency acoustic wave, 
$-{P_{\rm{amp}}}\sin (2\pi f_{\rm a}t)$, 
with the pressure amplitude $P_{\rm{amp}} = 125$ kPa and the frequency $f_{\rm a} = 29$ kHz. In the theoretical calculations, we set 
the polytropic exponent as $\gamma = 1.4$, and the saturated vapor pressure as $P_{\rm{v}} = 2338$ Pa (same for the subsequent all sections).
We measure the effect of gravity with the buoyancy parameter $\delta  = \sqrt {\rho g{R_{\rm{max}}}/{P_\infty }} $
in which 
$R_{\rm{max}}$ is the maximum bubble radius.
The effect of gravity is neglected since the buoyancy parameter $\delta$ is only 0.002. The theoretical and experimental results of the bubble radius variation are depicted in Fig. \ref{figure1}. Driven by the acoustic wave, the bubble expands to a maximum radius of 40 $\rm{\mu}$m before collapsing and continues to oscillate with a high frequency. The maximum radius after the first rebounding is merely 12 $\rm{\mu}$m, indicating a drastic loss in bubble energy. The above process is successfully predicted by our bubble theory and Gilmore's, Keller's, and Prosperetti's bubble equations, which can be proof of the validity of our theory for bubble dynamics in a free field neglecting gravity. Also, it is shown in Fig. \ref{figure1} that the result of the Rayleigh-Plesset equation deviates from the experiment after the first bubble oscillation because the model neglects fluid compressibility and thus cannot correctly reproduce the damping in maximum radii of the acoustic bubble. Note that the result obtained from the Gilmore equation deviates from other compressible models, which coincides with the analysis presented by \citet{pp86} that the accuracy of the Keller model is better than the Gilmore model.

\begin{figure*}
	\centering\includegraphics[width=0.8\linewidth]{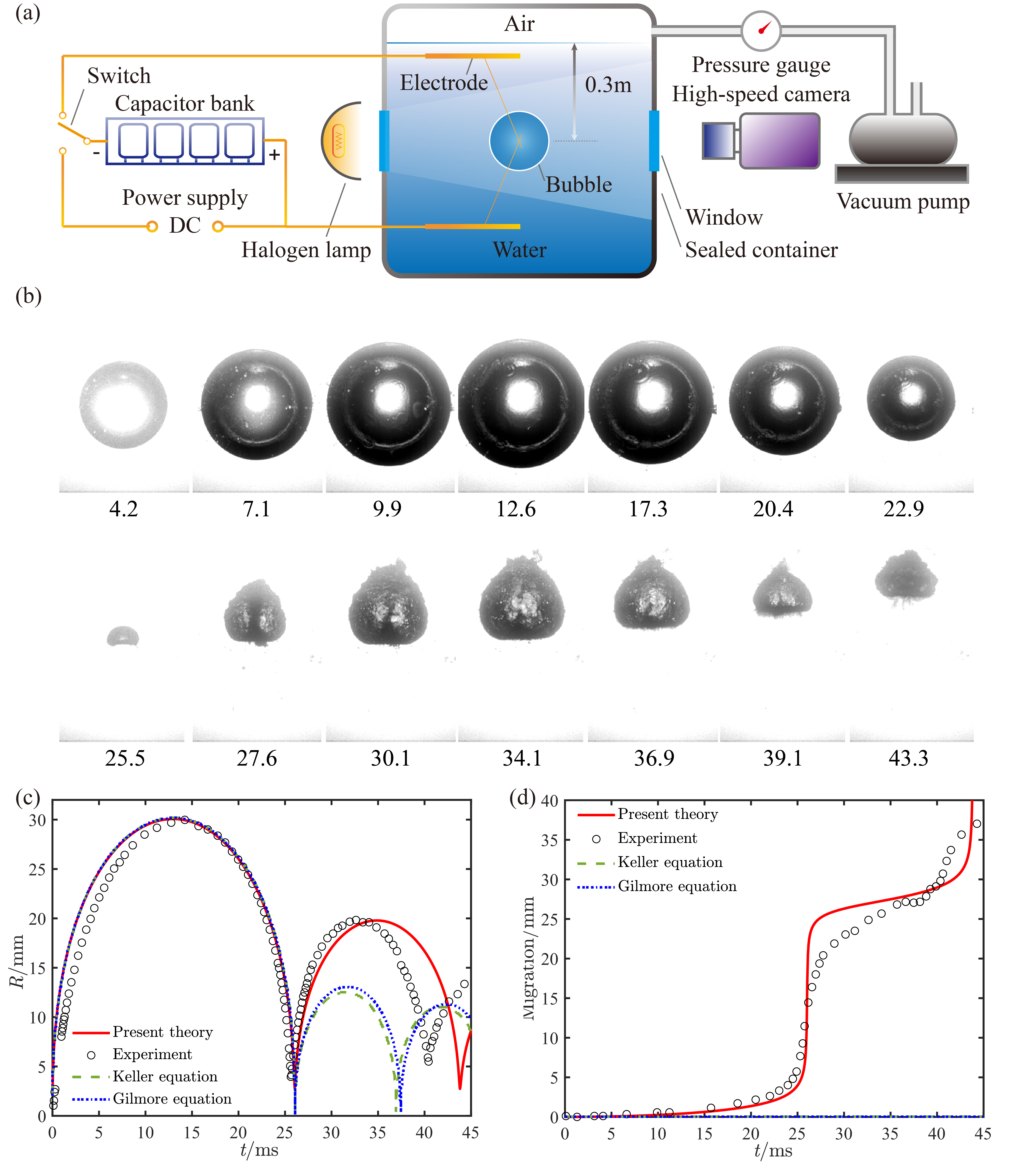}	
	\caption{{ Underwater spark-generated bubble experiment and comparisons of theoretical and experimental results in a gravity field.} {(\textit{a})} Schematic diagram of the experiment setup. {(\textit{b})} Selected sequential high-speed images of the spark-generated bubble. The capturing time is labeled below each image in milliseconds. Frame width, 51.62 mm. 
		{(\textit{c})} Comparisons between the theoretical and the experimental results of bubble radius. {(\textit{d})} Comparisons between the theoretical and the experimental results of the bubble migration.}
	\label{figure2}
\end{figure*}

\subsection{Bubble dynamics in a gravity field}
\label{S:11}

\begin{figure*}
	\centering\includegraphics[width=0.8\linewidth]{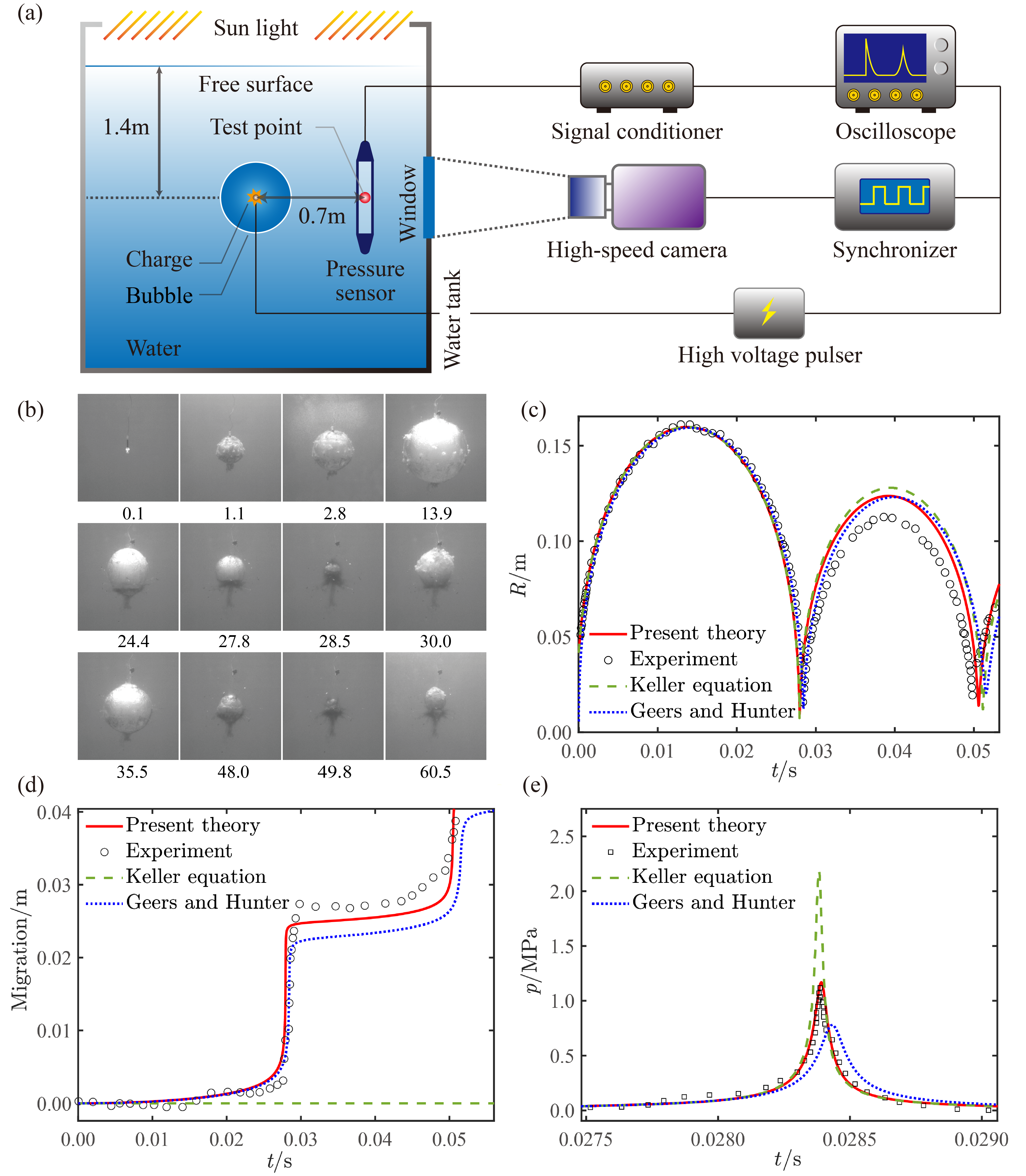}
	\caption{{ Underwater explosion bubble experiment and comparisons of theoretical and experimental results in a gravity field.}  {(\textit{a})} Schematic diagram of the experiment setup. {(\textit{b})} Selected sequential high-speed images of the underwater explosion bubble. The capturing time is labeled below each image in milliseconds. Frame width, 0.490 m. {(\textit{c})} Comparisons between the theoretical and the experimental results of the bubble radius in the gravity field. {(\textit{d})} Comparisons between the theoretical and the experimental results of the bubble migration in the gravity field. {(\textit{e})} Comparisons between the theoretical and the experimental results of the pressure pulse of the first collapse. The experimental pressure data is measured at the test point as in the schematic diagram. }
	\label{figure_undex_ff}
\end{figure*}

In this section, we continue to validate our theory for bubble dynamics in a gravity field using the experimental data of a spark bubble generated in a sealed container. The ambient pressure was reduced to 6.82 kPa to enhance the influence of gravity on the bubble behavior, resulting in significant bubble migration. The experimental setup is shown schematically in Fig. \ref{figure2}(\textit{a}) and further details can be found in \citet{zccw15}.
The bubble reached a maximum radius $R_{\rm{max}} = 29.8$ mm with the buoyancy parameter $\delta$ = 0.21. As shown in Fig. \ref{figure2}(\textit{b}), the bubble oscillated with significant upward migration due to gravity. 
For theoretical calculation, the initial conditions of the bubble are set as $P_0 = 45$ kPa, $ R_0 = 2.07 $ mm, and $ \dot{R}_0= 100$ m/s. In this paper, we determined the initial conditions for modeling spark and laser-induced bubbles with the present theory using the method introduced in \citet{w13} which
requires the maximum radius $R_{\rm{max}}$ and the minimum pressure $P_{\rm{min}}$ when the bubble radius reaches $R_{\rm{max}}$. $P_{\rm min}$ can be determined by the ratio $R_{\rm max2}/R_{\rm max}$ where $R_{\rm max2}$ is the maximum radius during the second oscillation cycle. A backward integration is performed starting from the bubble at the maximum expansion with zero wall velocity using the fourth-order Runge-Kutta method until the initial condition of the bubble is obtained.
The added mass coefficient and the drag coefficient in the bubble migration equation are set as 1.0 and 0.5, respectively. The theoretical results are plotted against the experiment in Fig. \ref{figure2}(\textit{c}) and (\textit{d}). 
The present theory yields a better prediction of the bubble radius variation compared to Keller and Gilmore's equations for the second bubble oscillation cycle.
It also correctly reproduces the strong bubble migration due to gravity which is not obtained with the other two models.
If we turn off the migration in our theory, the radius curve would be identical to that from the Keller equation. 
Thus, it is crucial to incorporate the effect of migration when buoyancy is significant.
There is still a deviation between the present results and the experiment mainly after the first oscillation cycle that could be possibly due to multiple factors, including the gravity-induced non-spherical bubble behavior, such as jetting/splitting, spark-induced combustion/plasma, and liquid/vapor phase transition during the process. In addition, more accurate results may be achievable by changing the EOS for the present theory. \\
\indent Next, we carry out further validation of the present theory with a small-charge underwater explosion bubble. The experiment setup is shown in Fig. \ref{figure_undex_ff}(\textit{a}). In the experiment, an explosive charge equivalent to 1.125 g PBX9501 (1.25 g TNT) was detonated at the center of a 4 m $\times$ 4 m $\times$ 4 m cubic tank filled with water. The water depth was adjusted so that the charge is 1.4 m below the free surface. The explosion generated a gas bubble with a maximum radius $R_{\rm{max}} = 0.16$ m. The effect of the water surface on the bubble is unsubstantial due to the distance to the surface being significantly larger than the maximum bubble radius. The buoyancy parameter is $\delta$ = 0.12. Due to the buoyancy effect, the bubble migrated upward while oscillating, as shown in Fig. \ref{figure_undex_ff}(\textit{b}). The initial conditions of the bubble are obtained with the method given in Appendix B as $R_0=42.52$ mm, $\dot{R}_0=61.80$ m/s, and $P_0=304.13$ kPa. The bubble dynamics are calculated using the present theory as well as the Keller equation and the Geers-Hunter model \citep{gh02}. The theoretical results are compared to the experimental data in Fig. \ref{figure_undex_ff}(\textit{c-e}). 
{The bubble radius variation predicted by our theory matches the experimental result with slightly higher accuracy than the other two models, as shown in \ref{figure_undex_ff}(\textit{c}). Also, the migration calculated with our theory is more comparable to the experiment than that with the Geers-Hunter model, see \ref{figure_undex_ff}(\textit{d}). The migration was not reproduced by the Keller equation. 
	In addition, Fig. \ref{figure_undex_ff}(\textit{e}) shows the comparison between the theoretical and the experimental results of the pressure pulse induced by the first bubble collapse. The pressure test point in the experiment is at the same depth as the charge with a distance of 0.7 m. }
Compared with the results from the experiment and the present theory, the Keller equation significantly overestimates the bubble pulse due to the absence of the migration, while the Geers-Hunter model underestimates it because of the introduction of excessive dissipation.

\subsection{Dynamics of multiple interacting bubbles}

\begin{figure*}
	\centering\includegraphics[width=0.8\linewidth]{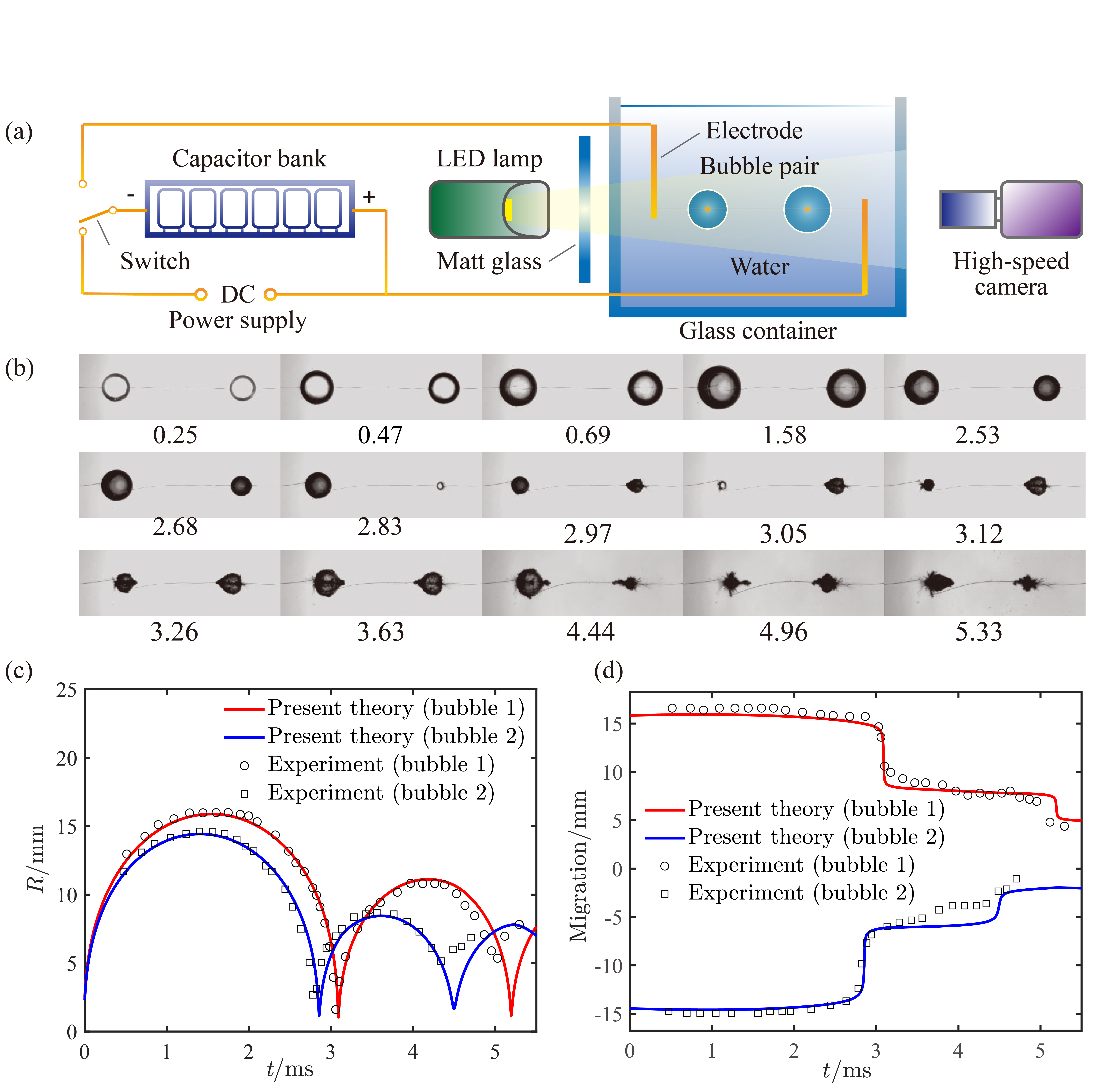}
	\caption{{The first experiment case of two interacting spark-induced bubbles and comparisons between theoretical and experimental results. (\textit{a}) Schematic diagram of the experiment setup. (\textit{b}) Selected sequential high-speed images of the spark-induced bubble pair. The capturing time is labeled below each image in milliseconds. Frame width, 149 mm. (\textit{c}) Comparisons between theoretical and experimental results of bubble radius variation. (\textit{d}) Comparisons between theoretical and experimental results of bubble migration.} }
	\label{f2bub}
	
\end{figure*}

In this section, we present experimental results of the dynamics of multiple interacting bubbles and further validated our theory by comparing the calculation to the experiment. In the first experiment, two bubbles were induced by electric sparks powered by a charged capacitor bank. We manually created two defects on a copper wire 0.21 mm in diameter that links the two poles of the capacitor bank. Upon discharge, two sparks were produced at the two defects due to increased local resistance, thus producing a bubble pair. Physical deviations in the two defects were intentionally created to give the bubbles differences in size and phase. The distance is 250 mm from the bubbles to the water surface and 200 - 250 mm to nearby walls, thus the bubbles are deemed far from boundaries considering the small radii (10 - 20 mm). A schematic diagram of the experiment setup is shown in Fig. \ref{f2bub}(\textit{a}). Two experiment cases were carried out where the two defects are horizontally placed at a distance of 93.6 mm and 42.7 mm, respectively, from each other, which are taken as the initial distances between the two bubbles in each pair. The bubbles are marked as bubbles 1 and 2 from left to right. In the first case, the maximum radii of the two bubbles were 16.0 mm and 14.6 mm and the first oscillation periods were 3.05 ms and 2.83 ms, respectively. 
The initial radii used for theoretical calculation are 
$R_{01}=2.50 \rm{}$ mm and
$R_{02}=2.29 \rm{}$ mm for bubble 1 and 2, respectively. 
The initial speeds of the bubble wall are
$\dot{R}_{01}=\dot{R}_{02}=130$ m/s and the initial pressure inside the bubbles are
$P_{01}=P_{02}=1.2 $ MPa.
The calculation is conducted from bubble initiation to the third oscillation period after which the experimental data of migration and radius become hardly measurable. The experiment and calculations for this case are summarized in Fig. \ref{f2bub}(\textit{b-d}). The high-speed images of the bubble pair are shown in (\textit{b}), where the two bubbles migrate towards each other during oscillation due to the Bjerknes force, which is similar to the attraction acting on an oscillating bubble by a nearby rigid wall. Time curves of the radius and the bubble center position are plotted in Fig. \ref{f2bub} (\textit{c}) and (\textit{d}), respectively. 
{
	The results calculated with our unified theory are compared to that of the experiments as in (\textit{c}) and (\textit{d}) and a good agreement is obtained.
}

\begin{figure*}
	\centering\includegraphics[width=0.8\linewidth]{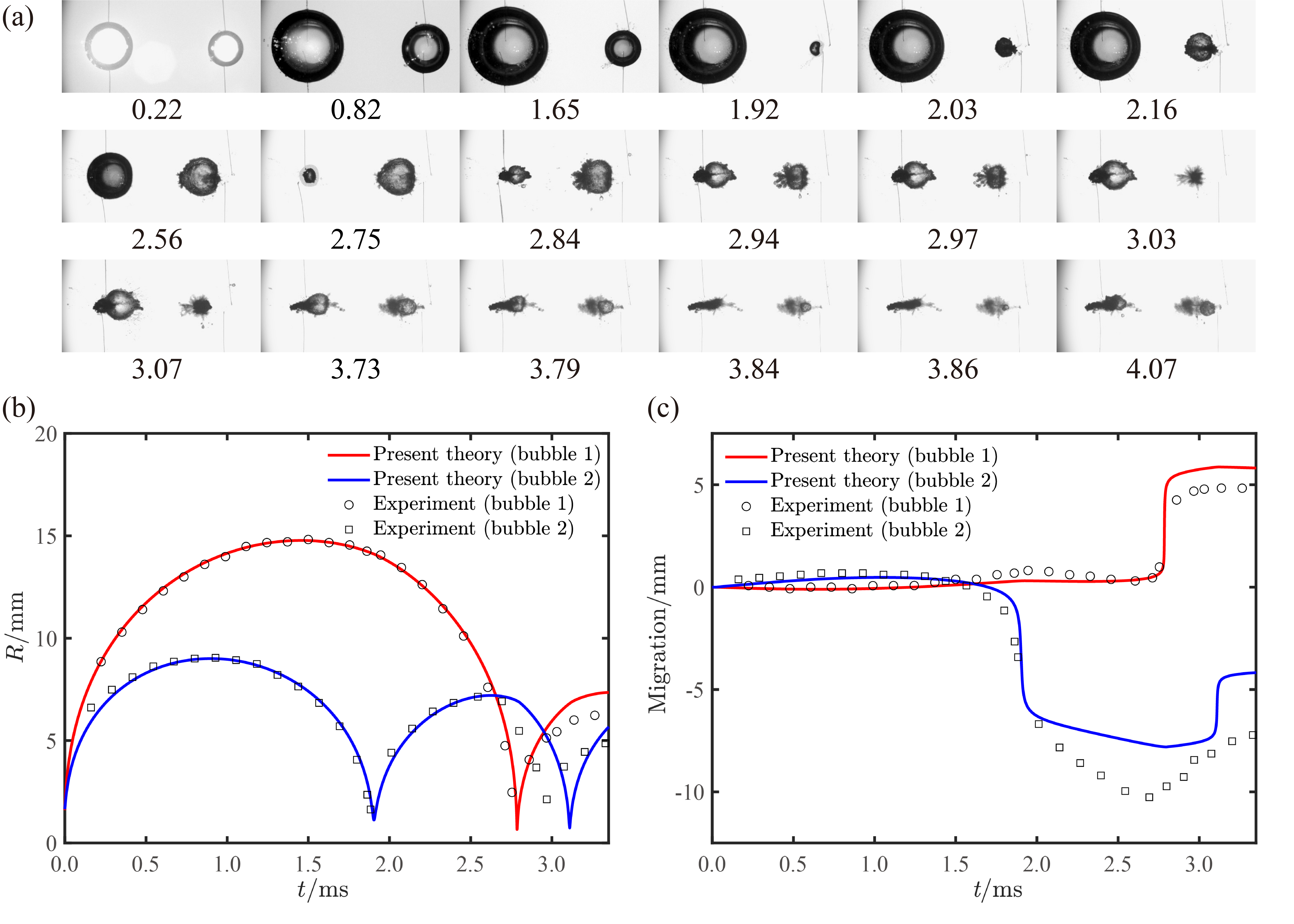}
	\caption{The second experiment case of two interacting spark-induced bubbles and comparisons between theoretical and experimental results. (\textit{a}) Selected sequential high-speed images of the spark-induced bubble pair. The capturing time is labeled below each image in milliseconds. Frame width, 96.86 mm. (\textit{b}) Comparisons between theoretical and experimental results of bubble radius variation. (\textit{c}) Comparisons between theoretical and experimental results of bubble migration. }
	\label{f2bub-2}
	
\end{figure*}

In the second experiment case, the energy difference between the two bubbles is increased. The maximum radii were 14.8 mm and 9.0 mm, respectively and the first oscillation periods were 2.92 ms and 1.75 ms, respectively. The high-speed images of the bubble pair are shown in Fig. \ref{f2bub-2} (\textit{a}).
The initial conditions for theoretical calculation are 
$R_{01}=1.60$ mm, 
$R_{02}=1.65$ mm,
$\dot{R}_{01}=240$ m/s,
$\dot{R}_{02}=110$ m/s,
$P_{01}=1.2 $ MPa, and
$P_{02}=1.2 $ MPa.
Due to the size difference, the two bubbles changes from being in-phase to out-of-phase over time. The effect of one bubble on its in-phase partner resembles that of a rigid wall, but the effect becomes similar to a free surface when the bubbles are out-of-phase. Thus, contrary to the previous case, the bubble migration was no longer monotonically towards each other. The center of bubble 2 reciprocated during multiple oscillation cycles, first migrating away from bubble 1 and then towards.
{
	The present theory is capable of reproducing such a phenomenon. 
	The prediction matches the experiment in bubble radius variation and captures the main features of the bubble migration, as shown in Fig. \ref{f2bub-2} (\textit{b}) and (\textit{c}). Possible reasons for the deviation may include the non-spherical bubble behavior, such as jetting, and the heat and mass transfer over multiple oscillation cycles.  
}




\begin{figure*}
	\centering\includegraphics[width=1.0\linewidth]{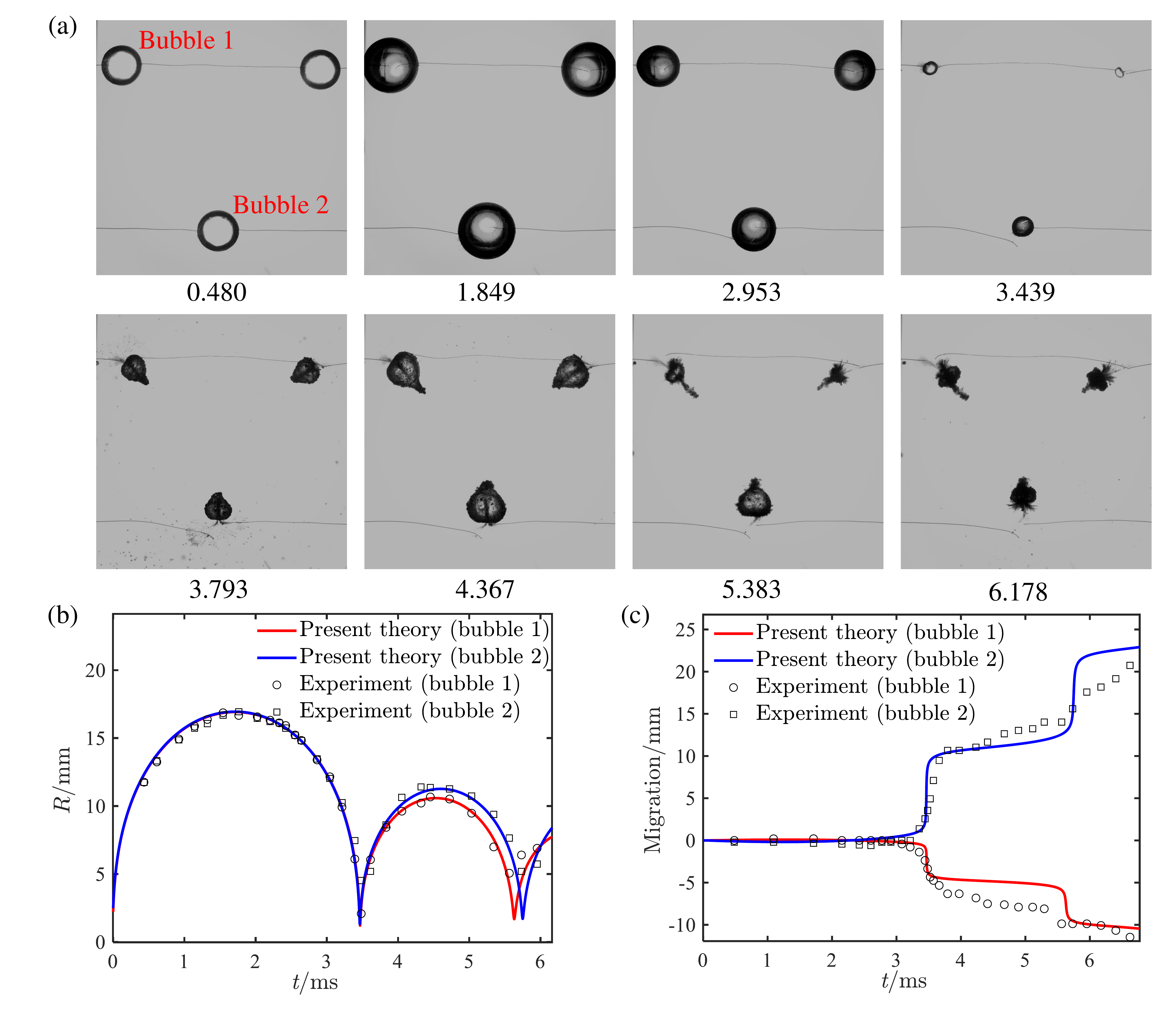}
	\caption{Comparison of the interaction of three spark-generated bubbles with the theoretical prediction. (\textit{a}) Selected sequential high-speed images of the three spark-induced bubbles. The capturing time is labeled below each image in milliseconds. Frame width, 147 mm. (\textit{b}) Comparison between theoretical and experimental results of bubble radius variation. (\textit{c}) Comparison between theoretical and experimental results of the bubble migration in the vertical direction.
	}
	\label{three-bubble}
\end{figure*}

Next, we carried out an experiment on the interaction of three electric spark-generated bubbles and compared the dynamics to theoretical results, as shown in Fig. \ref{three-bubble}, which introduces higher complexity. The three bubbles are positioned at the vertices of a triangle with the top two bubbles being 118 mm apart and the lower bubble 115 mm from the other two. 
Since the behavior of the upper two bubbles is almost symmetrical to each other, we only compare the upper left bubble and the lower one, marked as Bubble 1 and Bubble 2, respectively. The initial radii of bubbles 1 and 2 for theoretical calculations are ${R}_{01}=2.22$ mm and ${R}_{02}=2.48$ mm, and the initial oscillation speeds $\dot{R}_{01}=180$ m/s and $\dot{R}_{02}=150$ m/s, respectively. The initial internal pressure of both bubbles is 1.2 MPa. While oscillating in volume, the bubbles keep migrating towards each other, as shown in Fig. \ref{three-bubble}(a). The radius and the migration of the bubbles in the vertical direction in the experiment are well reproduced by the theoretical results, as shown in Fig. \ref{three-bubble}(b) and (c).

\begin{figure*}
	\centering\includegraphics[width=1.0\linewidth]{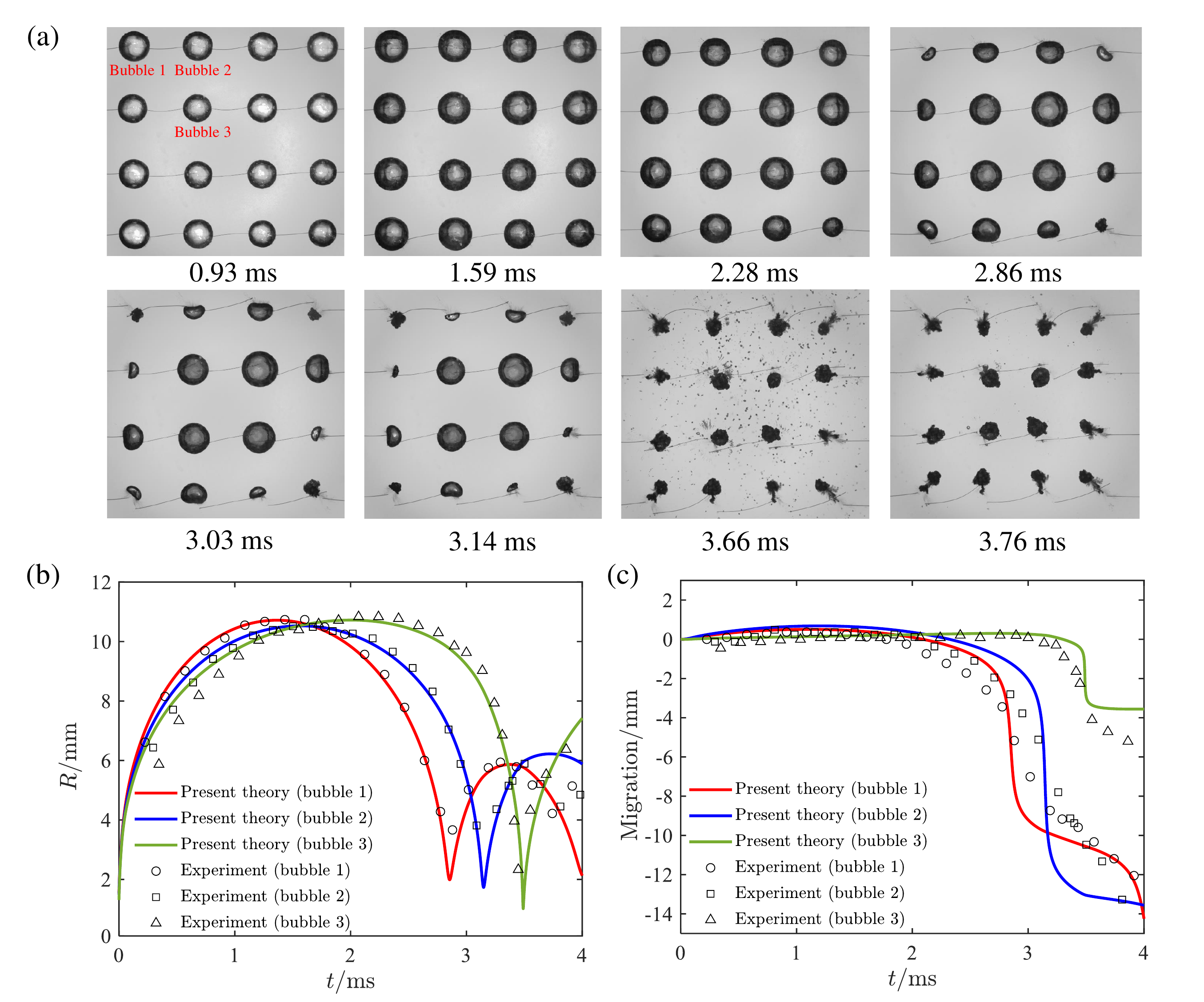}
	\caption{{\color{black}Comparison of the bubble dynamics in a cluster of sixteen spark-generated bubbles with the theoretical prediction. (\textit{a}) Selected sequential high-speed images of the bubble cluster. The black dots in the seventh frame are cavitation bubbles triggered by the pressure fluctuation due to the collapses and rebounds of the spark bubbles.  The capturing time is labeled below each image in milliseconds. Frame width, 151 mm. (\textit{b}) Comparison between theoretical and experimental results of the bubble radius variation. (\textit{c}) Comparison between theoretical and experimental results of the bubble migration in the vertical direction.}
	}
	\label{16bubble}
\end{figure*}

{\color{black}
	Further, the dynamics of a bubble cluster are modeled and compared with an experiment. The bubble cluster consisted of sixteen bubbles positioned uniformly on a 4 $\times$ 4 grid with an interval of 40 mm. All the bubbles were generated by the underwater discharge at the same instant. Considering the symmetry of the cluster, we only analyze the dynamics of three representative bubbles among them, marked as bubble 1, bubble 2, and bubble 3 as shown in Fig. \ref{16bubble}(a). 
	The bubble oscillation can be prolonged by neighbor bubbles with the same oscillation phase. Thus, bubble 3 surrounded by its neighbors from all 4 directions had the greatest oscillating period. On the contrary, bubble 1 located at the external corner was less affected by the other bubbles, which led to the shortest oscillation period. The bubbles away from the cluster center collapsed earlier and their re-expansion accelerated the final collapse of bubble 3. All the bubbles migrated towards the cluster center due to bubble interaction. 
	In the theoretical calculation, the initial internal pressure of all the bubbles is 1.2 MPa. The initial radii of the three selected bubbles are: ${R}_{01}=1.41$ mm, ${R}_{02}=1.36$ mm, and ${R}_{03}=1.18$ mm, respectively. The initial oscillation speeds are $\dot{R}_{01}=\dot{R}_{02}=180$ m/s and $\dot{R}_{03}=220$ m/s, respectively. To simplify theoretical calculations, the initial radii and speeds of other bubbles are identical to the three bubbles in a symmetrical manner, which can be considered as reasonable given the slight differences in bubble behavior in most of the first cycle.
	The radius, period, and vertical migration from the simulation with the present theory are in good agreement with the experimental results, as shown in Fig. \ref{16bubble}(b) and (c), demonstrating the theory's capability in dealing with complex bubble cluster dynamics.}

\subsection{Bubble dynamics near a boundary}
\label{S:12}

\begin{figure*}
	\centering\includegraphics[width=0.8\linewidth]{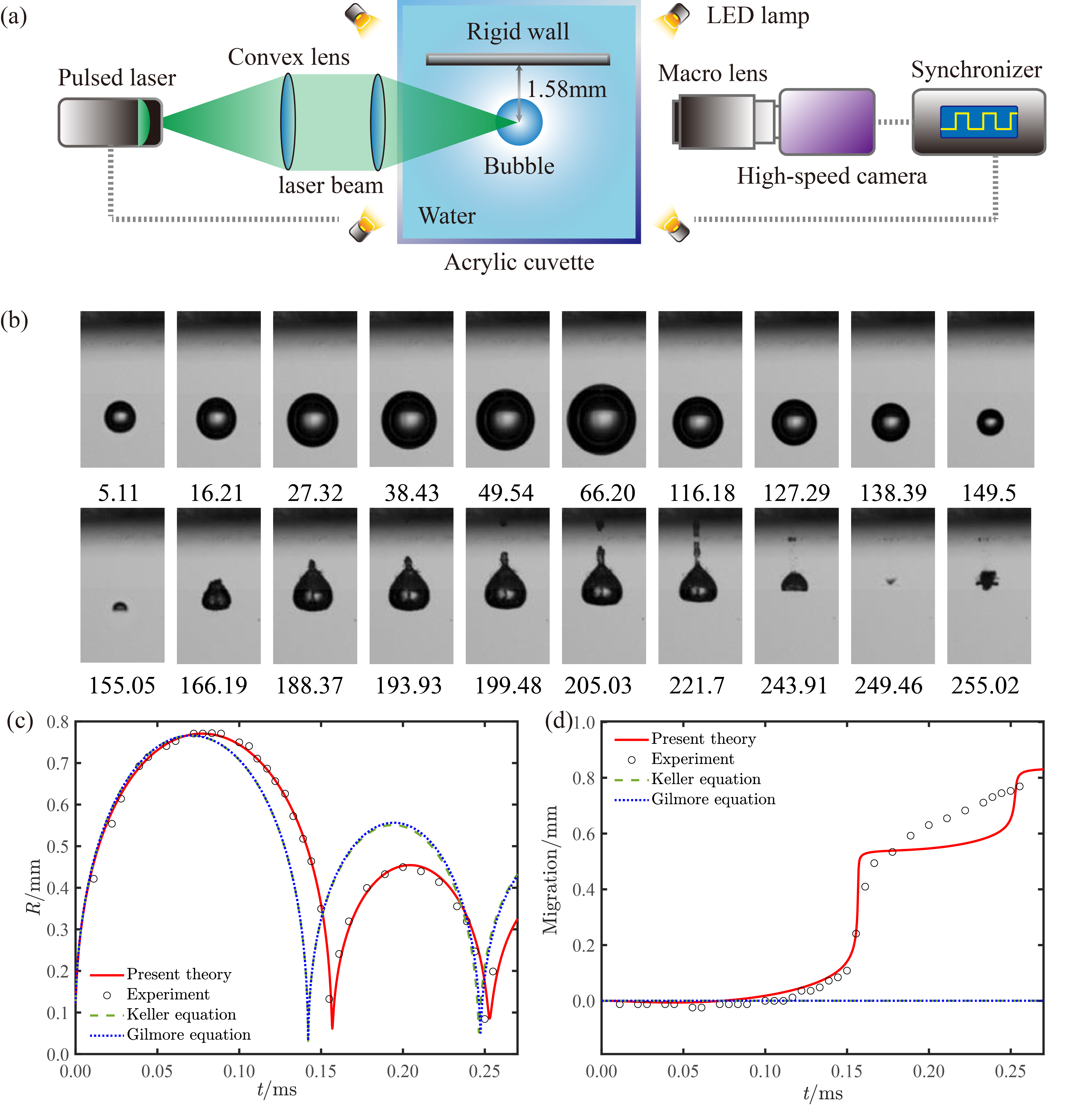}
	\caption{{Laser-induced cavitation bubble experiment near a rigid wall and comparisons between theoretical and experimental results.}  {(\textit{a})} Schematic diagram of the experiment setup. {(\textit{b})} Selected sequential high-speed images of the laser-induced cavitation bubble. The capturing time is labeled below each image in microseconds. Frame width, 2.16 mm. {(\textit{c})} Comparisons between  theoretical and  experimental results of  bubble radius variation. {(\textit{d})} Comparisons between  theoretical and  experimental results of  bubble migration.  }
	\label{figure5}
\end{figure*}

In this section, we validate our theory in the scenario of bubble oscillation near a rigid boundary. A cavitation bubble was generated at the atmospheric pressure by a Q-Switched Nd:YAG laser breakdown beneath a rigid boundary at a standoff distance of $d_\mathrm{b} = 2.02R_{\rm{max}}$, with the maximum bubble radius $R_{\rm{max}}$ = 0.768 mm. The experimental setup is shown in Fig. \ref{figure5}(\textit{a}) and high-speed images of the laser bubble oscillating while migrating towards the rigid boundary at the top are shown in Fig. \ref{figure5}(\textit{b}). For the theoretical calculation, we obtain the initial conditions of the bubble as $P_0 = 1.2$ MPa, $ R_0 = 0.121 $ mm, and $ \dot{R}_0= 130$ m/s  in the same way as stated before.
The results are plotted against the experimental data in Fig. \ref{figure5}(\textit{c}) and (\textit{d}). When a bubble oscillates near a rigid wall, the flow is retarded by the presence of the wall. Thus, the bubble oscillation period 
increases with a decreasing standoff parameter. 
The theories of Keller and Gilmore do not allow for the boundary effect, which may lead to a discrepancy in the oscillation period and the maximum radius of the second cycle when compared to the experiment. The present theory has considered the boundary effect and the calculated bubble radius and oscillation period are closer to the experimental data for multiple cycles, as shown in Fig. \ref{figure5}(\textit{c}). During the bubble-wall interaction, the pressure between the bubble and the wall is smaller than that on the opposite side, thus the bubble is pushed towards the wall, which is usually accompanied by a jet pointing to the wall. Our theory also takes into account the bubble migration induced by the boundary effect and the calculated migration curve is consistent with the experiment, as shown in Fig. \ref{figure5}(\textit{d}). {To further validate the present theory, we derive the Rayleigh time for a bubble oscillating near a rigid boundary by integrating the modified Rayleigh-Plesset equation
	\begin{equation}
	\label{RPwall}
	(R+\frac{R^2}{2{d_{\rm{b}}}}) \ddot{R} + \frac32 \dot{R}^2 (1+\frac{2 R}{3 {d_{\rm{b}}}}) = -\frac{P_{\rm{a}}}{\rho},
	\end{equation}
	in time, which is similar to \citep{r17,plesset49}. Then, the Rayleigh-like bubble oscillation period near a rigid wall can be expressed as
	\begin{equation}
	\label{periodwall}
	T = R_{\rm{max}}\sqrt {\frac{6 \rho}{P_{\rm{a}}}} \int\limits_0^1 {\sqrt {\frac{{{\eta^3}}}{{1 - {\eta^3}}}\left( {1 + \frac{\eta}{{{2d_{\rm{b}}}}}} \right)} d\eta},
	\end{equation}
	in which $\eta$ is the auxiliary variable. In this case, the Rayleigh-like period is 0.153 ms. Comparatively, the bubble period of the present theory is 0.156 ms, which is closer to the experimental result of 0.155 ms due to the consideration of multiple factors, e.g., the fluid compressibility, the internal pressure of the bubble, the surface tension, and the viscosity.}

\begin{figure*}   
	\centering\includegraphics[width=0.8\linewidth]{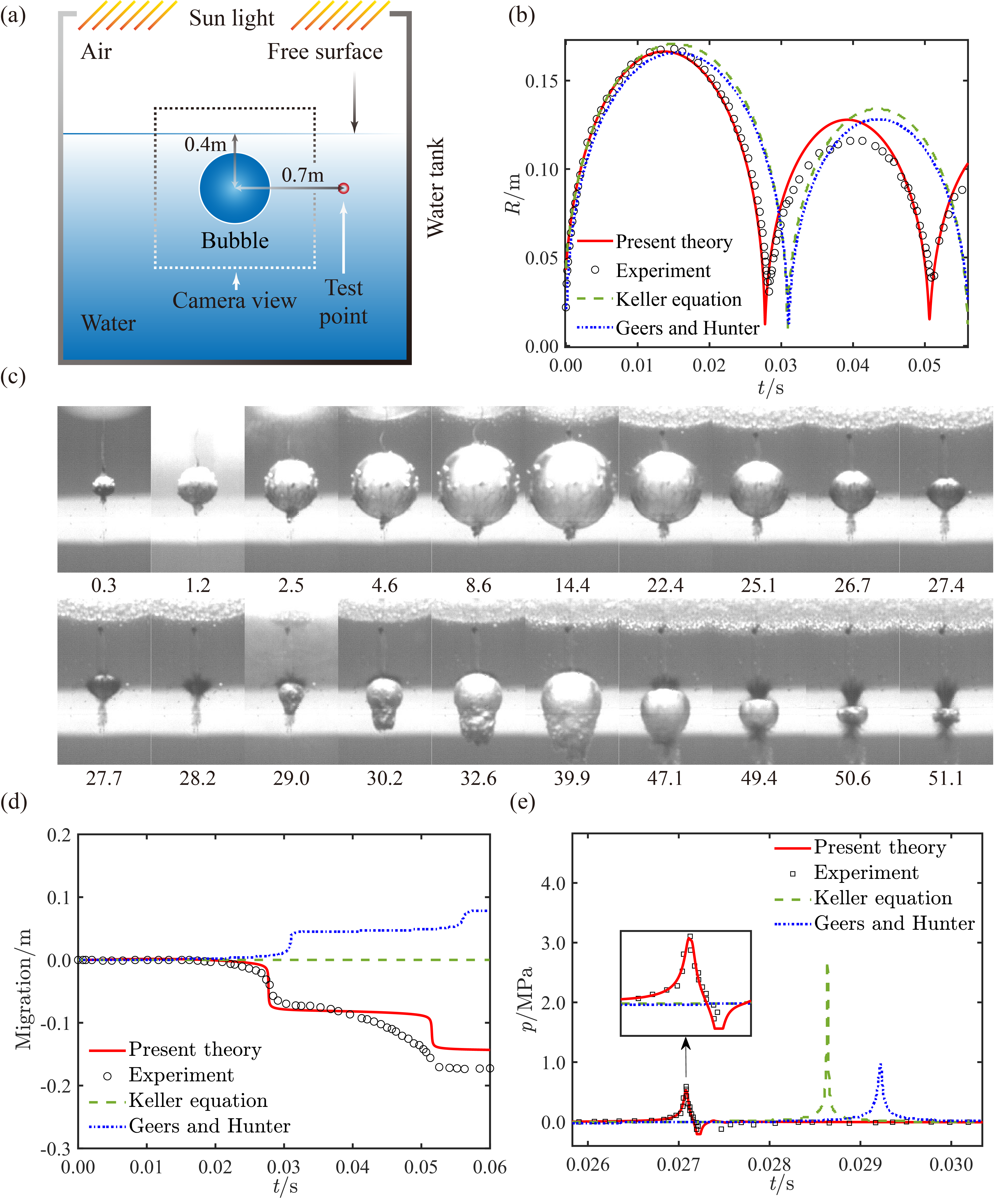}
	\caption{{Underwater explosion bubble experiment and comparisons of theoretical and experimental results near a free surface.}  {(\textit{a})} Schematic diagram of the experiment setup. {(\textit{b})} Selected sequential high-speed images of the underwater explosion bubble. The capturing time is labeled below each image in milliseconds. Frame width, 0.446 m. {(\textit{c})} Comparisons between the theoretical and the experimental results of the bubble radius considering gravity and boundary effect. {(\textit{d})} Comparisons between the theoretical and the experimental results of the bubble migration considering gravity and boundary effect. {(\textit{e})} Comparisons between the theoretical and the experimental results of the pressure pulse of the first collapse. The experimental pressure data is measured at the test point as in the schematic diagram.  }
	\label{figure6}
\end{figure*}

Additionally, we also compared the theoretical results of bubble dynamics near a free surface to the experimental data of a small-charge underwater explosion bubble. The experiment setup is shown schematically in Fig. \ref{figure6}(\textit{a}). In the experiment, an explosive charge equivalent to 1.125 g PBX9501 (1.3 g TNT) was detonated at the center of the aforementioned tank with a water depth of 0.4 m to generate a gas bubble with a maximum radius $R_{\rm{max}} = 0.169$ m. Due to the repelling effect of the free surface, the bubble migrated downward while oscillating, as shown in Fig. \ref{figure6}(\textit{c}).
The buoyancy parameter is $\delta = 0.126$. 
The initial conditions of the bubble are $R_0=43.57$ mm, $\dot{R}_0=61.83$ m/s, and $P_0=303.94$ kPa which are obtained with the method given in Appendix B.  
The bubble dynamics are calculated using different models and the results are compared to that of the experiments in Fig. \ref{figure6}(\textit{b}) and \ref{figure6}(\textit{d}). 
Compared to an infinite flow domain, the inertia of the liquid is reduced when a free surface is present, thus the bubble can oscillate faster and have a smaller period. The pressure of the bubble gas is much smaller than the atmospheric pressure during most of the lifetime of the bubble, thus the pressure gradient between the bubble and the free surface points upward and the bubble is repelled by the free surface, which is usually accompanied by a downward jet.

The Keller equation does not consider the boundary effect and bubble migration, while the Geers-Hunter model allows for migration but neglects the boundary effect. Consequently, there are deviations in the bubble radius and period to the experimental results, and a key feature that the bubble is repelled by the free surface is not reproduced. Compared to that, with the present model, the calculated bubble radius variation is in better agreement with the experiment and the migration of the bubble away from the free surface is also predicted. {Similar to equation (\ref{periodwall}), it is easy to find that the Rayleigh-like bubble pulsation period near the free surface can be expressed as
	\begin{equation}
	\label{periodfree}
	T = R_{\rm{max}}\sqrt {\frac{6 \rho}{P_{\rm{a}}}} \int\limits_0^1 {\sqrt {\frac{{{\eta^3}}}{{1 - {\eta^3}}}\left( {1 - \frac{\eta}{{{2d_{\rm{b}}}}}} \right)} d\eta}.
	\end{equation}
	Compared with the bubble periods (0.0276s and 0.0283s) calculated by equation (\ref{periodfree}) and from the experiment, the bubble period calculated by the present theory, 0.0277s, is also more accurate.}
In addition, Fig. \ref{figure6}(\textit{e}) depicts the comparison between the theoretical and the experimental results of the pressure pulse induced by the first collapse of the bubble. The pressure is measured at the same depth as the charge center and 0.7 m away in the horizontal direction. The pressure peak and the pulse width calculated by the present theory match the experimental data. The negative pressure induced by the rarefaction wave that is formed by the reflection of the pressure pulse at the free surface is also successfully captured by the present theory.

{\color{black} 
	We proceeded to compare the experimental data with the theoretical results obtained for multiple bubbles located near a free surface. In Fig. \ref{threefree}, we observe the progression of three bubbles in a linear arrangement near the free surface and compare the bubble radius and vertical migration between our theory and the experiment. At inception, all bubbles were positioned 47.6 mm away from the free surface. The left and right bubbles were initially located 52.8 mm and 55.8 mm away from the central bubble, respectively, which resulted in their almost symmetric configuration with respect to the central bubble. We focused our analysis on the dynamics of the left and central bubbles, which are identified as bubble 1 and bubble 2, respectively. Bubble 1 generated an oblique liquid jet and migrated in the same direction as bubble 2 under the combined action of the free surface and bubble 2. Bubble 2 experienced horizontal elongation during collapse, with both of its sides obstructed by the contracting bubbles on its left and right, while simultaneously migrating downwards, affected by the free surface. The initial conditions for bubbles 1 and 2 in the theoretical calculation were $R_{01}=2.92$ mm, $R_{02}=2.1$ mm, $\dot{R}_{01}=100$ m/s, $\dot{R}_{02}=180$ m/s, and $P_{01}=P_{02}=1.2$ MPa. The present theory was found to accurately reproduce the time histories of bubble radius and migration in the experiment, as presented in Fig. \ref{threefree}(b) and (c).
	\begin{figure*}   
		\centering\includegraphics[width=1.0\linewidth]{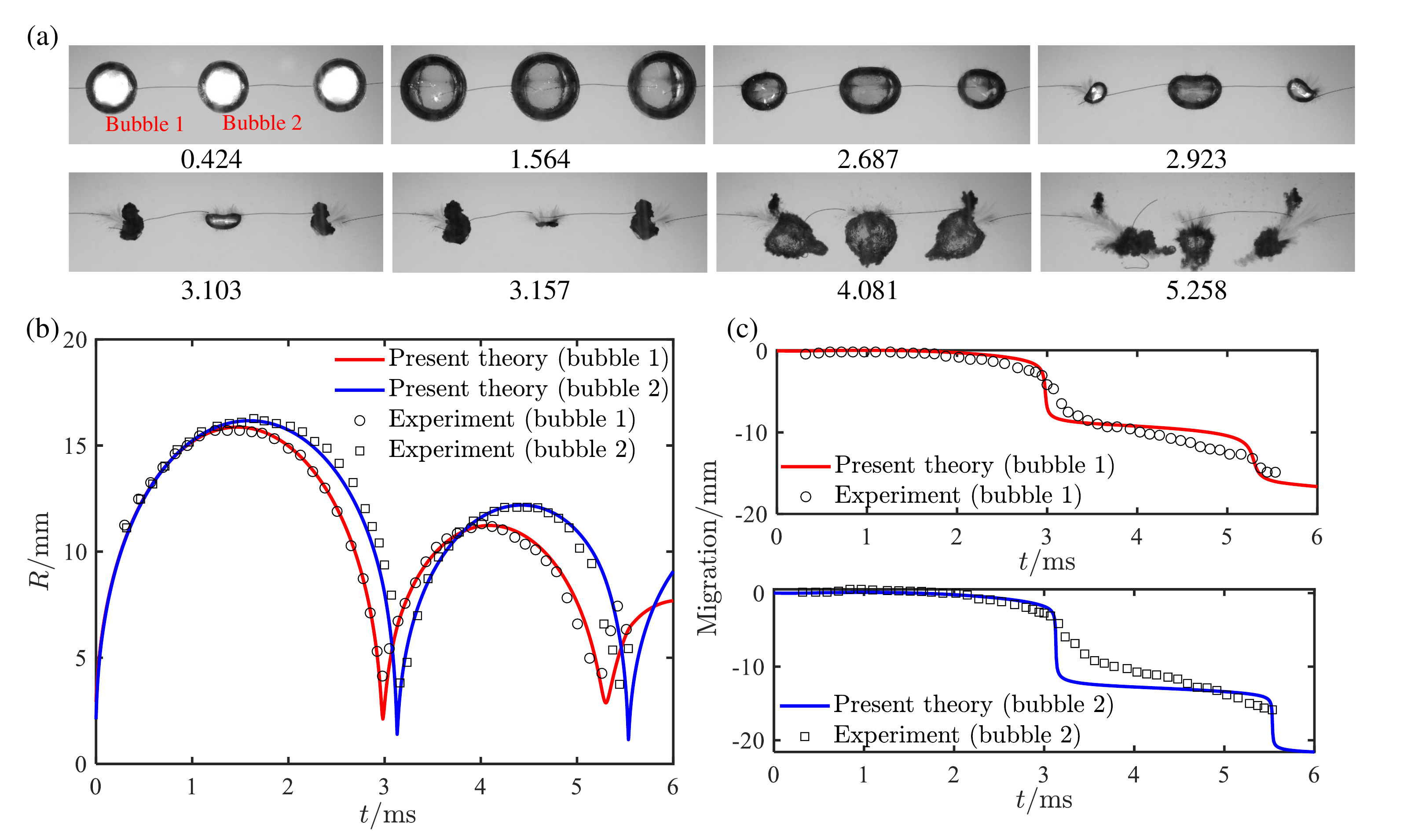}
		\caption{{\color{black} Comparison of the interaction of three spark-generated bubbles near the free surface with the theoretical results. (\textit{a}) Selected sequential high-speed images of the three bubbles. The capturing time is labeled below each image in milliseconds. Frame width, 153 mm. (\textit{b}) Comparison between theoretical and experimental results of bubble radius variation. (\textit{c}) Comparison between theoretical and experimental results of the bubble migration in the vertical direction.}  }
		\label{threefree}
	\end{figure*}
}

\begin{figure*}
	\centering\includegraphics[width=0.8\linewidth]{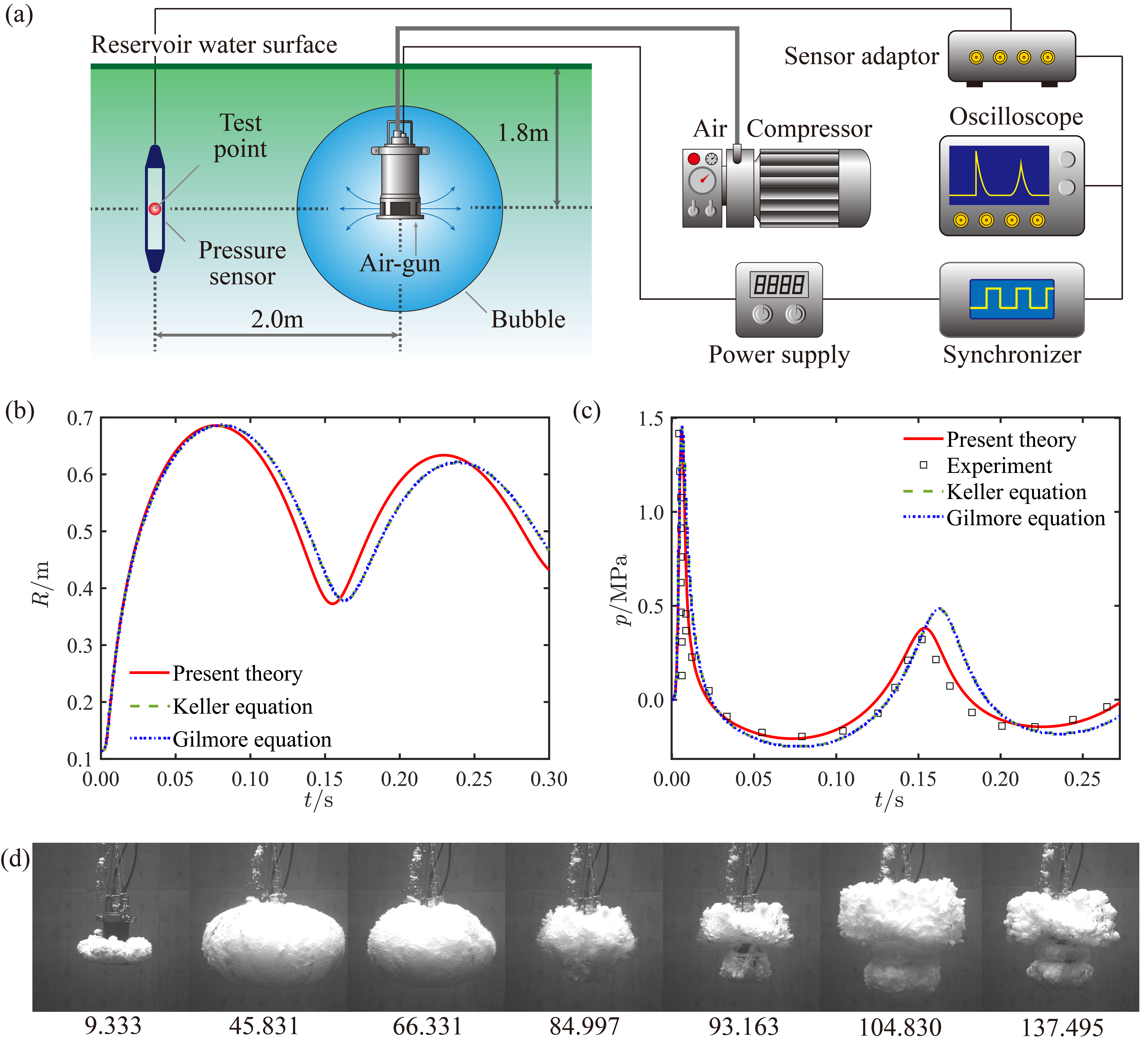}
	\caption{{Air-gun bubble experiment and comparisons of theoretical and experimental results near a free surface.  {(\textit{a})} Schematic diagram of the experiment setup. {(\textit{b})} Comparisons between the theoretical results of bubble radius variation versus time. {(\textit{c})} Comparisons between the theoretical and the experimental results of bubble-induced pressure variation versus time. {(\textit{d})} Selected sequential high-speed images of the air-gun bubble generated in a water tank with the initial internal pressure and volume of the chamber being $P_0=5$ MPa and $V_0=2458$ cm$^3$, respectively. The capturing time is labeled below each frame in milliseconds. Frame width, 113 cm.}}
	\label{figure_airgun}
\end{figure*}

In the last experimental verification, we applied our theory to model high-pressure air-gun bubble dynamics and compared the theoretical results against experimental data. When triggered, an air-gun rapidly released compressed air within a few milliseconds, forming a high-pressure bubble that expands and collapses. 
The experiments were conducted in a deep reservoir and a schematic diagram of the experiment setup is given in Fig. \ref{figure_airgun}{(\textit{a})}. An air-gun was placed 1.8 m below the water surface. 
The pressure in the flow field was measured by a pressure sensor at the same depth with a distance of 2.0 m from the air-gun. For the theoretical analysis, we use an air-release model to assess the mass flow from the air-gun chamber to the bubble and the ideal gas law to update the internal pressure of the bubble. Details of the implementation can be found in previous works \citep{shuai20}. 
Here, the initial internal pressure and volume of the chamber are $P_0=10$ MPa, $V_0=6227$ cm$^3$, respectively.
The results of the present theory, as well as that of the Keller equation and the Gilmore equation, are compared to the experimental data in Fig. \ref{figure_airgun}{(\textit{b})} and {(\textit{c})}. The present theory is capable of considering bubble migration and the effect of the free surface and, therefore, produces a result in better agreement with the experimentally obtained pressure profile. In addition, the above experiment was conducted in a reservoir where it was difficult to capture the bubble shape. Therefore, we present photos of an air-gun bubble produced in a water tank in Fig. \ref{figure_airgun}(d) where it was better observed, though certain boundary effect may exist.

\section{Discussion}\label{sect:discussion}
In this part, we discuss the applicability of our theory and then demonstrate its competence via implementation in the prediction of the complicated dynamic behavior of two interacting bubbles divergent in phase and energy, with new physics revealed.

\subsection{Applicability of the present theory}
\begin{figure*}
	\centering\includegraphics[width=0.8\linewidth]{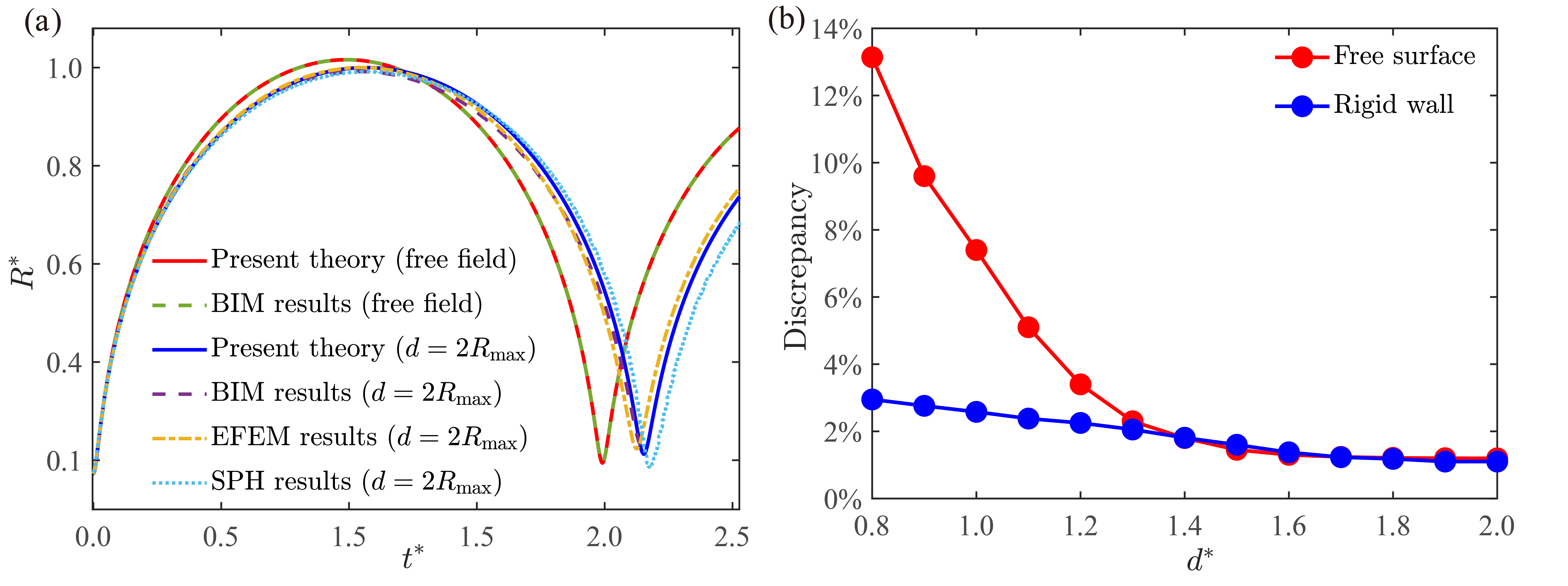}
	\caption{
		Comparisons between the theoretical result and the numerical simulation.
		{(\textit{a})}
		Bubble radius variations near a rigid boundary and in a free field obtained using the present theory and different numerical methods. 
		{(\textit{b})} 
		The discrepancy between the theoretical and numerical results of the first bubble oscillation period as a function of the standoff distance to the boundary scaled by $R_{\rm{max}}$.
	}
	\label{figure8}
\end{figure*}

Here, we analyze and discuss the applicability of the present 
theory in an extended parameter space through a quantitative comparison to numerical simulations. A number of numerical methods have been successfully applied to bubble dynamics, such as boundary integral method (BIM)/boundary element method (BEM) \citep{wb10,shuai20,Hanrui2021}, finite difference method (FDM)/finite volume method (FVM)/Eulerian finite element method (EFEM) \citep{tian2018analysis,wang2018bubble}, 
smooth partial hydrodynamics (SPH)
\citep{
	li2022improved}, and lattice-Boltzmann method (LBM) \citep{chang2017thermal,sofonea2018corner}. 
Here we choose three representative methods, namely  
BIM, EFEM, and SPH 
for the comparison. 
The effect of gravity, surface tension, and viscosity are ignored in both the theoretical and the numerical calculations to highlight the effect of boundaries on bubble dynamics. The initial conditions of the bubble are set as ${R_0} = 0.173{R_{\rm{max}}}$, ${P_0} = 100 P_{\rm{\infty}}$, $\gamma  = 1.4$, and $\dot R_0 = 0$. Fig. \ref{figure8}(\textit{a}) depicts the theoretical and the numerical results of bubble radius variations with the standoff distance to the rigid boundary being $d = 2R_{\rm{max}}$. Also, the results neglecting the boundary effect (i.e., in a free field) are presented. The boundary effect has a noticeable influence on the oscillation period of the bubble and the results of our theory are in good consistency with the numerical simulation. Hereafter, symbols with `*' are non-dimensional physical quantities with the length, pressure, and density scales being $R_{\rm{max}}$, $P_{\rm{\infty}}$ and $\rho$, respectively. The scales for the other quantities are calculated accordingly, e.g., $\sqrt{P_{\rm{\infty}}/\rho}$ for velocity and $R_\mathrm{max}\sqrt{\rho /P_{\rm{\infty}}}$ for time.

To further analyze the discrepancy between the two approaches quantitatively and evaluate the applicability of the present theory, we compare the results of the present theory to that of the BIM simulations within the first oscillation cycle of a bubble with varied standoff distances to a rigid boundary or a free surface, as depicted in Fig. \ref{figure8}(\textit{b}). 
{
	For scenarios with the rigid boundary, the discrepancy in period remains within 2\%, except for when the standoff distance is smaller than $1.3R_{\rm{max}}$, 
	For scenarios with the free surface, 
	the discrepancy is higher than that for the rigid boundary at small standoff distances possibly due to nonlinear free surface motions (such as a water spike). In spite of that, the discrepancy drops to about 2\% as soon as the standoff distance increases to $1.4R_{\rm{max}}$.
}

\subsection{Dynamics of interacting bubbles}

\begin{figure*}
	\centering
	\includegraphics[width=0.6\linewidth]{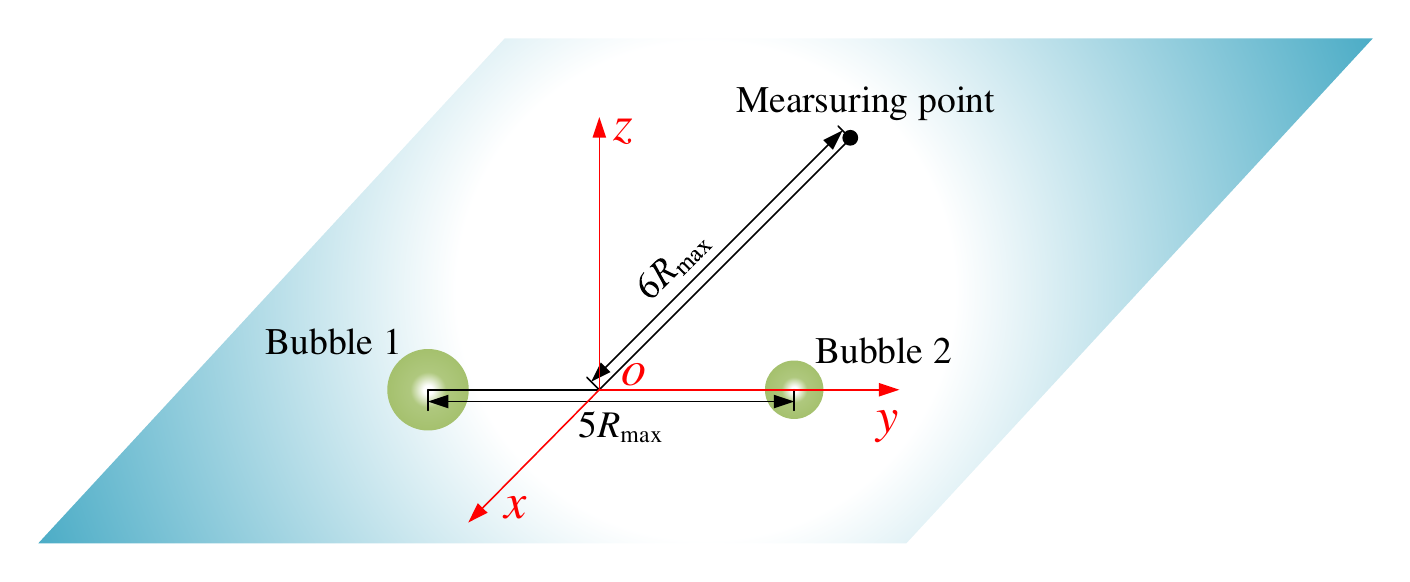}
	\caption{Sketch for the configuration of the interacting bubble pair and the measuring point for fluid field pressure.}
	\label{fig:sketch-for-double-bubble}
\end{figure*}

Next, we further apply the present 
theory to complex scenarios of bubble dynamics to reveal new underlying physics. The interaction between two bubbles is fundamental to the study concerning the dynamics of multiple bubbles which is a problem frequently encountered in a wide range of applications and troubles the researchers with sophisticated variations in bubble behavior influenced by many factors. Therefore, we have chosen the dynamics of a two-bubble system with different initial bubble energy ratios and initiation times as the objective. A sketch of the two-bubble system is shown in Fig. \ref{fig:sketch-for-double-bubble}. Herein, we take the maximum radius of bubble 1 as the length scale and ignore the gravity. The Mach number $Ma = \sqrt{P_{\rm{\infty}}/\rho}/C$ is taken as 0.013. The initial conditions of  the two bubbles are given by the same $P_ 0 = 100 P_{\rm{\infty}}$ and $\dot{R}_0 = 0$. The initial radius of bubble 1 is $R_0 = 0.175R_{\mathrm{max}}$ and that of bubble 2 is adjusted to tune the energy ratio between the two bubbles.  
The parameters considered are defined as follows. The internal energy of a bubble at initiation is estimated as $E_i=P_{0,i}V_{0,i}/(\gamma-1)$ where $i=1$ or 2 indicates physical quantities of bubble 1 or 2, respectively. We then denote $E^*=E_1/E_2$ as the bubble energy ratio. The initiation of bubble 1 is marked as $t=0$. The phase difference between the bubbles is defined as $\beta=\Delta t^*/T_{\mathrm{f}}^*$ where $\Delta t^*$ is the non-dimensional initiation time of bubble 2 and $T_{\mathrm{f}}^*$ is the non-dimensional first oscillation period of bubble 1 in a free field without bubble 2.

\begin{figure*}
	\centering\includegraphics[width=0.86\linewidth]{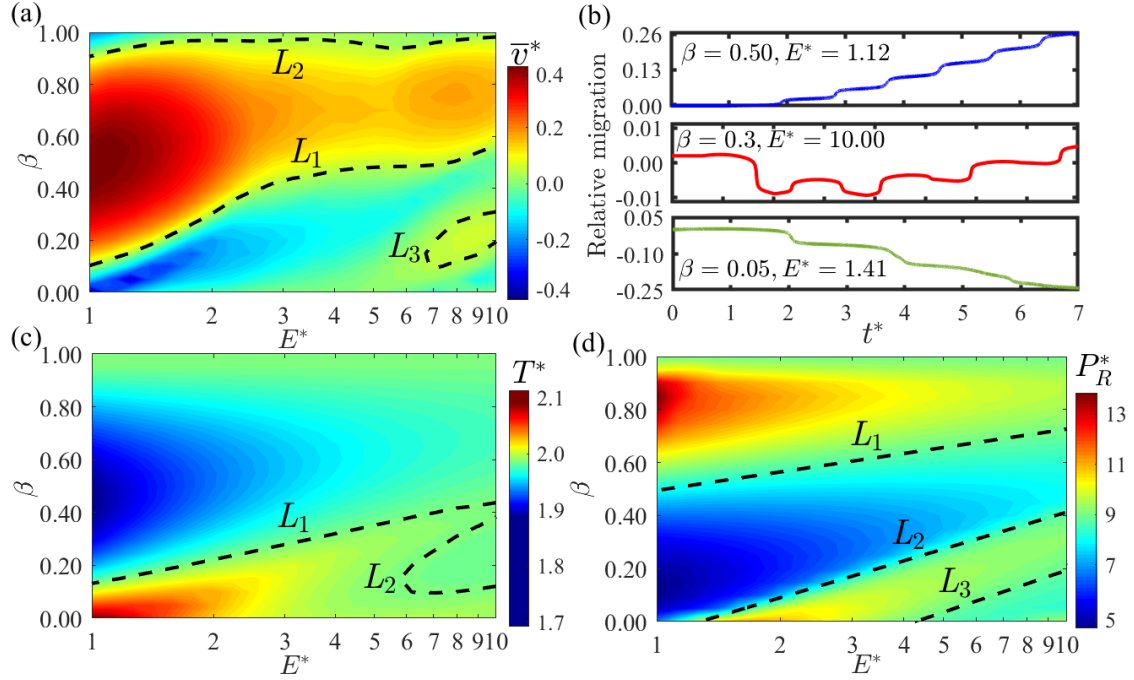}
	\caption{{ Dynamics of two interacting bubbles with varied initial energy ratio $E^*$ and phase difference $\beta$.} {(\textit{a})} Variation of bubble migration, represented by the relative speed between the two bubbles averaged over the first three and a half oscillation periods of bubble 1. The positive values indicate bubbles repelling each other and the negative attracting. L1, L2, and L3 represent where the relative speed is zero. {(\textit{b})} Typical time curves of the relative migration between the two bubble centers. {(\textit{c})} Variation of bubble period, represented by the first oscillation period of bubble 1 subjected to interaction with bubble 2. On L1 and L2, the value equals the first period of bubble 1 in a free field without other bubbles. {(\textit{d})} Variation of the bubble-induced pressure peak at a unit distance, indicated by the highest pressure peak reached inside bubble 1 multiplied by its radius. On L1, L2, and L3, the value equals that of bubble 1 in a free field without a second bubble.
	}
	\label{Contour1}
\end{figure*}

The parameter space is set as follows. The energy ratio $E^*$ ranges from 1 to 10 and the phase difference $\beta$ from 0 to 1. The initial distance between the two bubbles is set as a constant (5 times the $R_{\mathrm{max}}$ of bubble 1 in a free field).
We focus on three properties that reflect the main bubble dynamics, namely, the migration, the oscillation period, and the scaled internal pressure. The results are presented in Fig. \ref{Contour1} in an $E^*-\beta$ parameter space. The migration is represented by the average relative speed of the two bubbles as depicted in Fig. \ref{Contour1}(\textit{a}). The positive values represent the two bubbles eventually repelling each other and negative values represent attracting, with three demarcation lines $L_1-L_3$. The isoline $L_1$ indicates that, at $E^*=1 $ where the two bubbles are initiated with the same energy, the migration alters from attracting to repelling when $\beta$ exceeds approximately 0.1. The threshold in $\beta$ gradually grows with $E^*$ and reaches approximately 0.55 at $E^*=10$, which means that a larger phase difference is required for the bubbles to repel each other when the bubble energy difference increases. $L_2$ resides within $0.9<\beta<1.0$ for different energy ratios, indicating a transition from repelling to attracting when the initiation of bubble 2 is delayed until the end of the first contraction stage of bubble 1. In the region indicated by $L_3$ where $E^*$ exceeds 6.6 and $\beta$ is around 0.1--0.3, the migration turns from attracting to repelling again. In fact, in an extended area containing $L_3$, the bubble-center distance reciprocates around its initial value during bubble oscillation. An example of the corresponding complex relative migration--time curve is depicted in Fig. \ref{Contour1}(\textit{b}). An explanation is that with the large energy difference, the migration of bubble 2 is dominated by the surrounding fluid motion set into oscillation by bubble 1. In addition, the scenarios with the highest repelling amplitude occur around $\beta=0.5$ and $E^*=1$. Here, the effect of bubble 1 on bubble 2 is similar to a free surface and thus the latter migrates away from the former. The scenarios with the highest attraction amplitude occur around $\beta=0$ and $E^*=1$, i.e., the two-bubble system is symmetrical. Here, the dynamics of a bubble mimics that near a rigid boundary, and hence the migration towards each other is prominent.

The variation of the non-dimensional first period of bubble 1, $T^*$, is shown in Fig. \ref{Contour1}(\textit{c}) which reflects the average energy of bubble 1 in the period in the ambient flow induced by bubble 2. On the isolines $L_1$ and $L_2$, $T^*$ equals the first period of bubble 1 in a free field without other bubbles. The pattern of Fig. \ref{Contour1}(\textit{c}) is similar to Fig. \ref{Contour1}(\textit{a}) but with opposite signs. The effects of bubble 2 on the pulsation period of bubble 1 can also be interpreted by analogy with the effects of a boundary. $T^*$ reaches a minimum of 1.877 near $E^*=1$, $\beta=0.45$, in which case bubble 2 initiates as bubble 1 approaches the maximum volume. In this scenario, the expansion of bubble 2 accelerates the contraction of bubble 1, similar to the effect of a free surface. From here, increasing $E^*$ leads to higher $T^*$ due to the weaker influence of bubble 2 with lower energy. Compared to $E^*$, altering $\beta $ results in more obvious changes in $T^*$. As $\beta$ approaches 1.0, $T^*$ grows until reaching $T_{\mathrm{f}}^*=1.986$; as $\beta$ approaches 0, isoline $L_1$ is crossed and $T^*$ exceeds $T_{\mathrm{f}}^*$
by a maximum of 5 \%. This is because the bubbles are in-phase and hinder each other, leading to a prolonged period similar to the bubble pulsation near a rigid boundary. In the region around $L_2$, $T^*$ varies in a very limited range and approximates $T_{\mathrm{f}}^*$, since bubble 2 hardly affects the period of bubble 1 when $E^*$ is large.



The scaled internal bubble pressure is reflected in Fig. \ref{Contour1}(\textit{d}) by a nephogram of $P_{\mathrm{R}}^* = P_{\mathrm{max}}^* R^*$, where $P_{\mathrm{max}}^*$ is the non-dimensional maximum pressure inside bubble 1 and $R^*$ the non-dimensional radius. $P_{\mathrm{R}}^*$ equals the pressure peak a bubble induces at a unit distance and implies the energy of bubble 1 at its first minimum volume. The three demarcation lines are the isolines for $P_{\mathrm{R}}^*$ being equal to what bubble 1 induces in a free field, $P_{\mathrm{Rf}}^*$, when bubble 2 is absent. If bubble 1 is expanding when the pressure wave emitted by bubble 2 arrives, the pressure wave does negative work to bubble 1 and decelerates the expansion, in which case a smaller $P_{\mathrm{R}}^*$ is expected. By contrast, if bubble 1 is collapsing, the pressure wave will accelerate the collapse and $P_{\mathrm{R}}^*$ will increase. The energy transmission rate increases with $R^2\dot{R}$, which is measured on bubble 1 at the incidence of the pressure pulse induced by bubble 2. The negative value indicates the transmission from bubble 2 to bubble 1, while the positive value indicates the reversed transmission.
The variation in $P_{\mathrm{R}}^*$ with $\beta$ reduces with the increase in $E^*$, since a higher energy ratio means a smaller size of bubble 2 and lower influence on bubble 1. For the region above $L_1$, $P_{\mathrm{R}}^*$ remains greater than $P_{\mathrm{Rf}}^*$. This indicates energy transmission from bubble 2 to bubble 1. 
The greatest $P_{\mathrm{R}}^*$, i.e. the highest transmission, occurs in a region around $\beta=0.85$, $E^*=1$, with $P_{\mathrm{R}}^*= 1.56P_{\mathrm{Rf}}^*$. This is because that the highest $-R^2\dot{R}$ is found right in the above region. 
Therefore, the expansion of bubble 2 intensifies the collapse of bubble 1, leading to the high $P_{\mathrm{R}}^*$. The required minimum $\beta$ for $P_{\mathrm{R}}^*>  P_{\mathrm{Rf}}^*$ starts as about 0.5 at $E^*=1$ and increases almost linearly until reaching about 0.75 at $E^*=10$, as reflected by the demarcation line $L_1$.
For the region below $L_1$, the pressure wave of bubble 2 arrives at bubble 1 at its expansion phase and does negative work to it. Thus, $P_{\mathrm{R}}^*$ falls below $P_{\mathrm{Rf}}^*$, indicating energy transmission from bubble 1 to bubble 2. The minimum $P_{\mathrm{R}}^*$ is about 4.5 and appears around $\beta = 0.1$ and $E^* = 1$ when $R^2\dot{R}$ arrives at the maximum. 
Additionally, it's worth noticing that as $\beta$ approaches 0, $P_{\mathrm{R}}^*$ may also exceed   $P_{\mathrm{Rf}}^*$ for certain ranges of $E^*$, corresponding to the region between $L_2$ and $L_3$. Here, the collapse of bubble 2 occurs at the late contraction stage of bubble 1. The pressure wave induced by the collapse of bubble 2 serves as an energy input to bubble 1 and that explains why $P_{\mathrm{R}}^*$ exceeds $P_{\mathrm{Rf}}^*$.

\begin{figure*}
	\centering\includegraphics[width=0.8\linewidth]{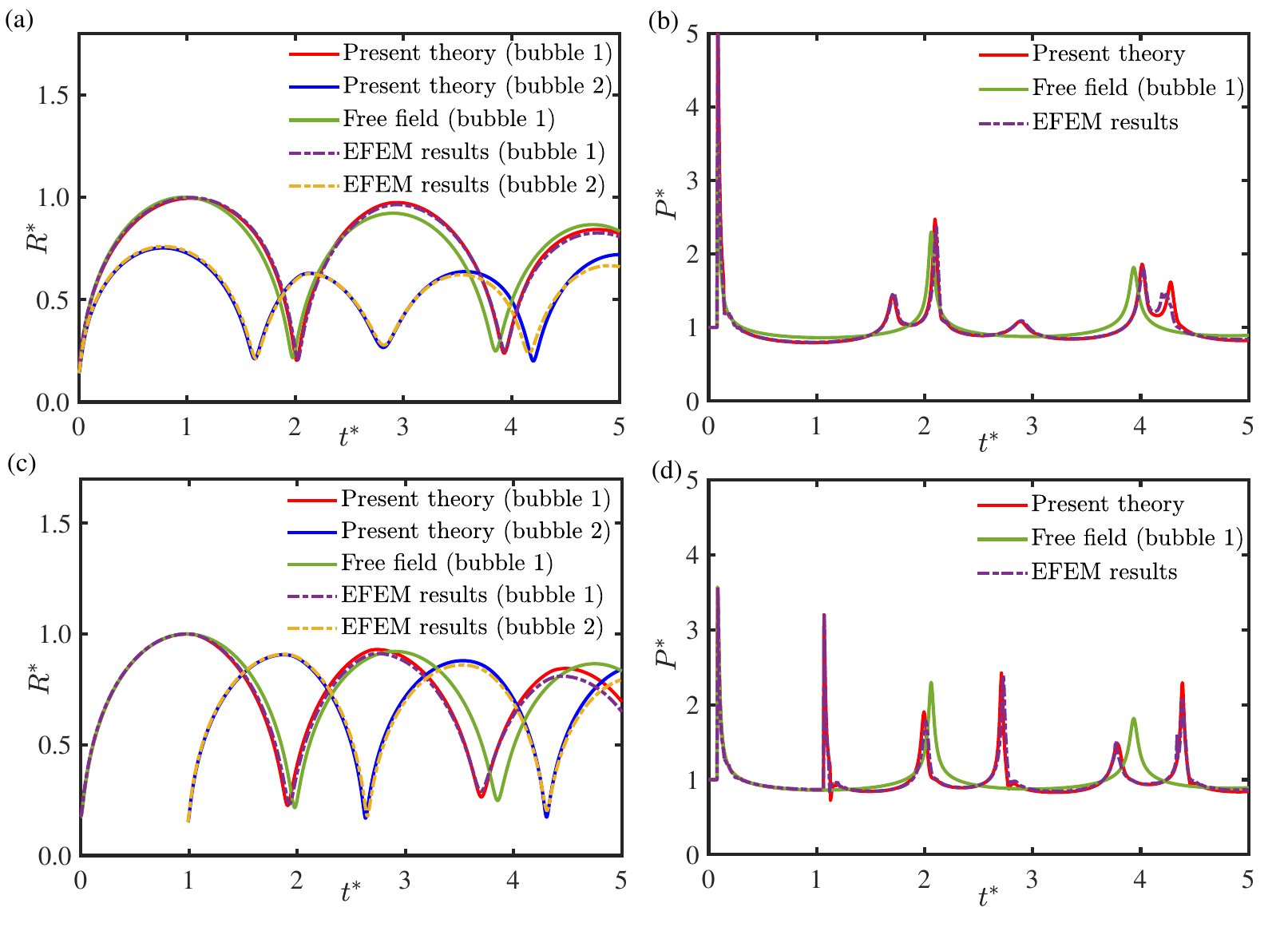}
	\caption{
		{Theoretical and numerical results of the dynamics of two interacting bubbles with phase and energy differences obtained using the present unified theory and the Eulerian Finite Element Method (EFEM), respectively.
			{(\textit{a})} Temporal variation of bubble radii for the bubble pair with $\beta$ = 0 and $E$ = 2 (the first scenario). 
			{(\textit{b})} Temporal variation of the pressure induced by the bubble pair in the flow field with $\beta$ = 0 and $E$ = 2. 
			{(\textit{c})} Temporal variation of bubble radii for the bubble pair with $\beta$ = 0.5 and $E$ = 1.6 (the second scenario). 
			{(\textit{d})} Temporal variation of the pressure induced by the bubble pair in the flow field with $\beta$ = 0.5 and $E$ = 1.6. 
		}
	}
	\label{figure10}
\end{figure*}

To further discuss and demonstrate the capability of the present theory in the prediction of complex pressure waves, energy transition, and complex physical laws of bubble dynamics, we have chosen two typical scenarios from the above-mentioned parameter space, i.e., ($\beta, E$) = (0, 2) and (0.5, 1.6), and modeled the corresponding bubble dynamics using the present theory. In the first scenario, the bubble inceptions are simultaneous, while in the second, bubble 2 is produced when bubble 1 is at the maximum expansion. We also compared the results to that of the numerical simulation using the EFEM. The radius variation and the bubble-induced pressure profiles are compared to the results of numerical simulations using the EFEM in \ref{figure10}. The test point where the pressure profiles are measured is indicated in Fig. \ref{fig:sketch-for-double-bubble} which is at the same depth as the bubble pair with a non-dimensional distance of 6.0 from the center of the configuration. 
The results of a single-bubble scenario where bubble 2 is removed are also added in the figure to highlight the influence of bubble 2. 

The radius variation and pressure profiles for the first scenario with the simultaneously generated bubbles are given in Fig. \ref{figure10}(\textit{a}) and (\textit{b}), respectively.
One may find that the maximum radius of bubble 1 in its second oscillation cycle has increased when bubble 2 is present. Using the bubble energy formulas 
\citep{KhooWang2004} 
which is based on the maximum bubble radius, we calculated the dimensionless energy of the two bubbles and find that the maximum energy of bubble 1 increased significantly from 4.12 when oscillating alone to 4.68 when interacting with bubble 2. It indicates that bubble 1, in its first two oscillation cycles, absorbs the energy of bubble 2 
during the bubble interaction. This is also reflected in the pressure profile, i.e., the first bubble period is prolonged, and the bubble collapse pressure is increased than that when bubble 2 is absent. 
Interestingly, the maximum radius of bubble 2 in different cycles also fluctuates. The maximum radius in the fourth cycle is noticeably higher than those in the second and the third. The dimensionless energy of bubble 2 in the fourth cycle reached 1.99, which is significantly higher than 1.54 and 1.55 of the two preceding cycles,  indicating a gain in the energy of bubble 2, rather than simply damping. The energy gain is from the positive work done by the pressure waves induced by the collapse of bubble 1 occurring in the third and the fourth cycle of bubble 2. This is also reflected in the pressure profile, i.e., the pressure peak induced by the third collapse of bubble 2 is remarkably higher than that of the previous two collapses and comparable to that of the second collapse of bubble 1. 
The above indicates complex energy transmission between the two interacting bubbles. 

The results of the second scenario where bubble 2 is generated when bubble 1 reaches its maximum expansion is shown in Fig. \ref{figure10}(\textit{c}) and (\textit{d}).
The maximum radii for both bubbles are damping during oscillations. The peaks on the pressure profile fluctuate, but the peaks induced by the same bubble upon each collapse keep reducing. 
Intriguingly, a negative pressure peak is captured at about $t^*$ = 1 in the theoretical result. It corresponds to the rarefaction wave generated after the pressure pulse induced by the initial expansion
of bubble 2 reflecting on the surface of bubble 1. Such ghost reflections are also manifested in the subsequent process of the bubble interaction. In general, good agreement is found between the theoretical and the numerical (EFEM) results of the bubble radius variation and the complex bubble-induced pressure fluctuation in the flow field of both scenarios, which proves the capability of the present theory in explaining the sophisticated bubble dynamics.

\section{Conclusion}\label{sect:conclusion}

Initiating from basic mathematical principles and physical equations, we established a novel and extendable
theory for 
oscillating bubble dynamics. 
It can simultaneously consider the complex physical factors including boundaries, bubble interaction, ambient flow field, gravity, bubble migration, fluid compressibility, viscosity, and surface tension. At the same time, it retains a unified and elegant mathematical form when dealing with various bubble-related problems. The theory achieves the unification of classical bubble theories such as the Rayleigh-Plesset equation, the Keller-Miksis equation, and the Gilmore equation.
We systematically compared the results of the present theory with those of the previous ones, the experiments, and the numerical simulations of bubble dynamics with a variety in bubble scales, sources, boundaries, and ambient conditions. It shows that the present theory has higher 
accuracy and applicability
compared to the classical ones. It can accurately predict the key bubble dynamics features including bubble oscillation, migration, and collapse-induced pressure pulses under different conditions and shows potential in exploring more sophisticated physics and mechanisms behind bubble dynamics such as bubble interaction, coupling of bubble-induced pressure waves, and inter-bubble energy transfer. 
The theory presented in this work may 
provide new references for future explorations in bubble dynamics, cavitation, and other multiphase flow-related problems.

\begin{acknowledgments}
	This work is funded by the National Natural Science Foundation of China (51925904, 52088102), the National Key R\&D Program of China (2022YFC2803500, 2018YFC0308900), Finance Science and Technology Project of Hainan Province (ZDKJ2021020), the Heilongjiang Provincial Natural Science Foundation of China (YQ2022E017) and the Xplore Prize. 
\end{acknowledgments}



\section*{\textsf{Conflict of Interest}} 
The authors declare no competing interests. 

\section*{\textsf{Data Availability}} 
All data needed to evaluate the conclusions are presented in the paper. The source codes used in the paper can be found at  {\color{blue}https://github.com/fslab-heu/unified-bubble-theory.git}.

\appendix

\section{Solution of a moving singularity for wave equation}
Assuming that there is a source at the origin with a strength of $f_0(t)$, the solution of the wave equation can be written as 
\begin{equation} \varphi_{f_0}(\boldsymbol{r},t) = \frac{1}{|\boldsymbol{r}|} f_0(t-|\boldsymbol{r}|/C).
\end{equation} 
Making use of the linearity of the wave equation, the solution of the wave equation considering the movement of the source can be obtained by superposition of a set of $\phi_{f_0}$ as
\begin{equation}\label{eq-source-moving}
{\color{black}\varphi_f(\boldsymbol{r},t) = 
\frac{C}{\left(C-\boldsymbol{v}\cdot\boldsymbol{r}_t\right)|\boldsymbol{r}_t|} f(t-|\boldsymbol{r}_t|/C)}
\end{equation}
where $\boldsymbol{r}_t$ is the vector pointing from the location of the source at $t-|\boldsymbol{r}_t|/C$ to $\boldsymbol{r}$. It is not easy to obtain an explicit analytic expression for $\boldsymbol{r}_t$ unless the source is moving with a constant velocity. Once equation (\ref*{eq-source-moving}) is obtained, the solution for a moving dipole with a strength of $q(t)$ can be calculated subsequently with the following limit
\begin{equation}\label{eq-dipole-moving}
\varphi_q = \lim_{L \rightarrow0}\frac{1}{2L}\left(\varphi_f(\boldsymbol{r}+\boldsymbol{e}L,t)-\varphi_f(\boldsymbol{r}-\boldsymbol{e}L,t)\right).
\end{equation}

\section{Initial conditions for underwater explosion bubble}
\subsection{Bubble formation and near-field shock wave propagation}\label{sect-near-field}
In this part, we present an approach to model the initial phase of an underwater explosion bubble. 
Upon the formation of the explosion bubble, the fluid pressure around the bubble is very high and a strong shock wave emits from the bubble. The basic assumption of the wave equation is thus violated. 
Therefore, we solve the Euler equation in the conservation-law form to treat this phase until the pressure and velocity are low enough for the initial condition of the unified theory for bubble dynamics.
Assume that there is a gas bubble located in still water starting to expand due to high internal pressure. The thermal conduction and viscous effect are ignored. 
The fluid motion is described by the 1-dimensional spherically symmetric Euler equations: 
\begin{equation}
\frac{\partial \rho}{\partial t} + \frac{\partial \rho u }{\partial r} = - \frac{2u}{r} \rho,
\end{equation}
\begin{equation}
\frac{\partial  \rho u}{\partial t} + \frac{\partial \rho u^2 +p }{\partial r} = - \frac{2u^2}{r} \rho,
\end{equation}
\begin{equation}
\frac{\partial E}{\partial t} + \frac{\partial u(E+p)}{\partial r} = 
-\frac{2u}{r}(E+p),
\end{equation}
where $E = \rho e + \frac12 \rho u^2$ is the total energy and $e$ is the specific internal energy per unit mass.
The right-hand sides of the above equations represent the geometric source terms in a spherical symmetric coordinate system. 

\begin{figure*}
	\centering
	\includegraphics[scale=1.8]{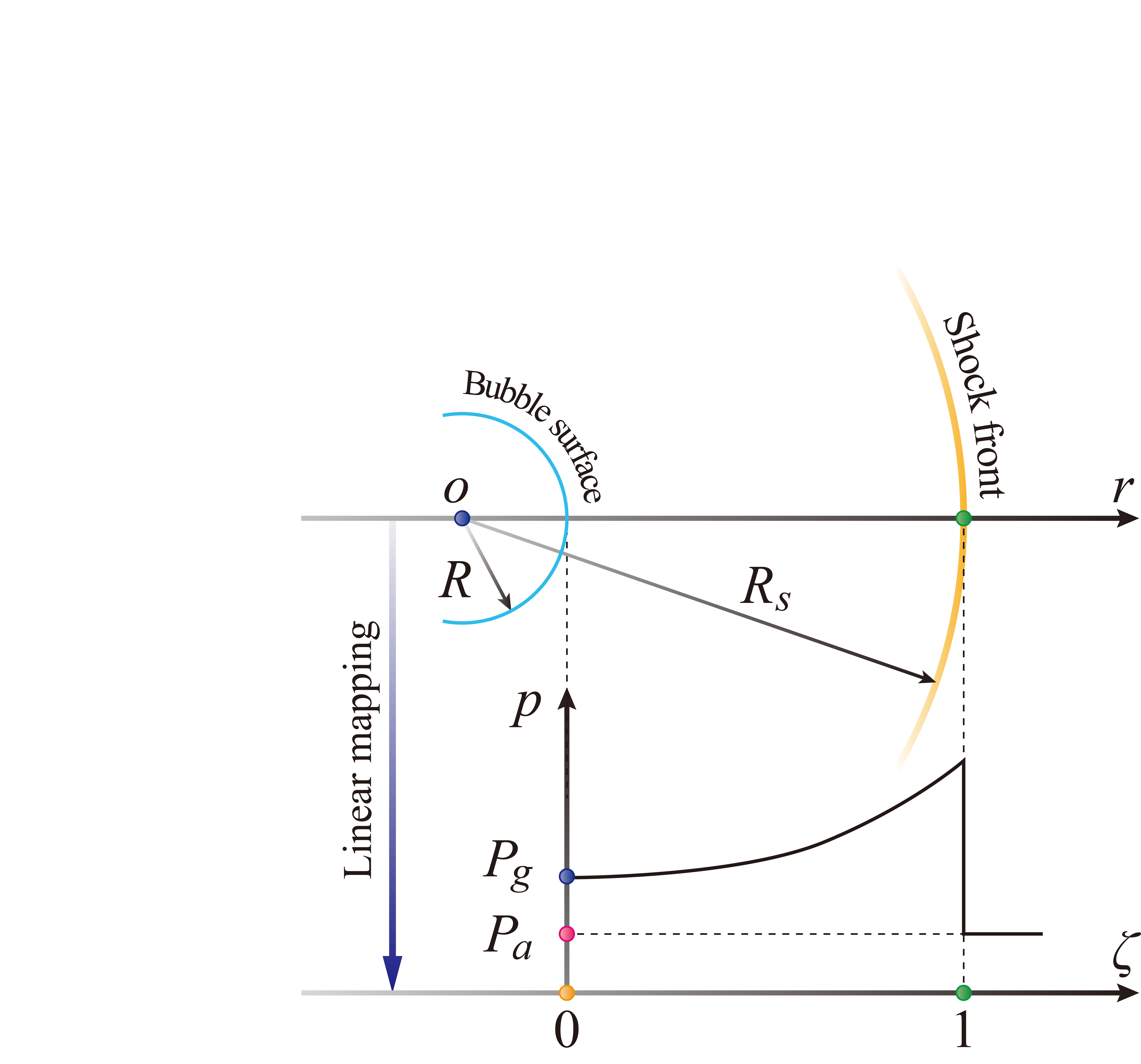}
	\caption{{Sketch for the linear mapping between the physical domain ($r$) to the numerical solution domain ($\zeta$) for the near-field solver.
		}
	}
	\label{fig:dg-shockwave}
\end{figure*}

After the detonation, a shock wave forms at the explosive charge surface and propagates into the water. The gaseous product of the explosive forms a bubble that expands rapidly. The radii of the shock front and the bubble surface are denoted as $R_\mathrm{s}$ and $R$, respectively. Thus, at $r>R_\mathrm{s}$, the fluid is undisturbed and at $r<R$ is the gaseous product inside the bubble. These two parts provide the boundary conditions of the fluid within $R<r<R_\mathrm{s}$. In this work, we focus on the fluid between $R$ and $R_\mathrm{s}$ that is smooth and solvable with high-order methods and avoid the difficulties in treating the multi-medium interface and the discontinuity at the shock front.
The physical coordinate $r$ is mapped linearly into the coordinate for the numerical solution $\zeta$ as
\begin{equation}
\zeta = \frac{r-R}{L},
\end{equation}
which is also illustrated in Fig. \ref{fig:dg-shockwave},
where $ L = R_\mathrm{s} - R $ is the length of the computational domain. 
Then we have
\begin{equation}
u = \frac{dr}{dt} = 
\frac{d\zeta}{dt}L+\zeta \dot{L}+\dot{R}.
\end{equation}
Denote $\hat{u} = \frac{d\zeta}{dt}$ as the fluid velocity observed in the coordinate system of $\zeta$ .  Thus, 
\begin{equation}
\hat{u} = \frac{1}{L}(u - \zeta \dot{L} - \dot{R} ).
\end{equation}

Denote $\hat{\partial}/\hat{\partial} t = {\partial}/{\partial} t + (\zeta\dot{L}+\dot{R}){\partial}/{\partial} r $ as the derivative with respect to $t$ when $\zeta$ is fixed. Then, the Euler equations can be rearranged and expressed in the following compact form:
\begin{equation}\label{eq-shockwave-near}
\frac{\hat{\partial}\boldsymbol{U}}{\hat{\partial}t} + 
\frac{{\partial}\boldsymbol{F}}{{\partial}\zeta} 
= \boldsymbol{S},
\end{equation}
where
\begin{equation}
\boldsymbol{U} = \begin{bmatrix}
\rho \\ \rho u \\ E
\end{bmatrix};
\boldsymbol{F} = \begin{bmatrix}
\hat{u}\rho  \\ \hat{u}\rho u + p/L \\ \hat{u}E+up/L
\end{bmatrix};
\boldsymbol{S} = -\frac{2}{r}\begin{bmatrix}
u\rho \\ u^2 \rho  \\ u(E+p)
\end{bmatrix} - \boldsymbol{U}\frac{\dot{L}}{L}.
\end{equation}
Note that an extra source term emerges in the scaled coordinate system.
Then, the hydrodynamic pressure at $r$ can be given by
\begin{equation}
p(r,t) = \begin{cases}
P_\infty &\text{ when } r>R_\mathrm{s}, \\
P_\mathrm{g} &\text{ when } r<R, \\
p_{l}{(\zeta,t)} &\text{ for else.}
\end{cases}
\end{equation}
Here, $p_l$ is the pressure calculated with the equation of state of the surrounding fluid.

The Euler equation system is closed by the equation of states of the fluids. Typically, the internal gas produced by explosive is modeled with the JWL equation\citep{JWL_EOS}, which reads
\begin{equation}
P_\mathrm{g} = A \exp\left(-R_1 \frac{R^3}{R_{\mathrm{c}}^3}\right)+
B\exp\left(-R_2 \frac{R^3}{R_{\mathrm{c}}^3}\right) + C \left(\frac{R}{R_{\mathrm{c}}}\right) ^{-3(\omega +1 )},
\end{equation}
where $R_{\mathrm{c}}$  is the radius of the explosive charge and $A, B, R_1, R_2$, and $ \omega$ are material related constants.

If the surrounding fluid is liquid, we model it by the Tammann EoS
\begin{equation}
p = \rho e (\gamma-1)-\gamma P_w.
\end{equation}

The initial conditions are calculated as follows. Initially, it is assumed that a high-pressure gas bubble is placed in a still fluid field. The internal gas is considered as still at the beginning for simplicity. Thus, a Riemann problem is formulated at the interface between the internal gas and the surrounding fluid. Here, we define the starting time ($t=0$) of the explosion bubble dynamics as when the detonation wave reaches the explosive charge surface. We assume that the charge has fully converted into gaseous products and thus formed a bubble at this moment. We start the simulation from a short time $ \epsilon$ after it. The fluid quantities are assumed to be constant between the bubble surface and the shock front. At $t = \epsilon$, the fluid pressure of the water between $R$ and $R_\mathrm{s}$  equals the inner pressure of the bubble, i.e., $P_\mathrm{g}$. The Mach number for the shock wave $M = \dot{R}_s/C_\infty$ is calculated with the Huginiot condition 
\begin{equation}
P_\mathrm{g}+P_w = \frac{P_\infty+P_w}{\gamma+1}(2\gamma M^2 -\gamma +1) .
\end{equation}
Then, the initial conditions for the fluid density and velocity are given by
\begin{equation}
\rho{(\zeta,\epsilon)} = \frac{\rho_\infty(\gamma+1)M^2}{(\gamma-1)M^2+2},
\end{equation}
and 
\begin{equation}
u_{(\zeta,\epsilon)} = \frac{2C_\infty}{\gamma+1}\frac{M^2-1}{M},
\end{equation}
respectively. Here, the subscript $\infty$ indicates the physical quantities of the undisturbed water.
The initial bubble radius $R$ and the shock front radius $R_\mathrm{s}$ at $t=\epsilon$ are given by 
\begin{equation}
\begin{cases}
R = R_{\mathrm{c}}+\epsilon u{(\zeta,\epsilon)}, \\
R_\mathrm{s} = R_{\mathrm{c}}+\epsilon M C_\infty,
\end{cases}
\end{equation}
where $R_{\mathrm{c}}$ is the radius of the explosive charge. In this paper, the value of $\epsilon$ is determined by letting $L = R_{\mathrm{c}}/100$ at $t=\epsilon$.

Within this framework, we can obtain the boundary conditions. For the boundary representing the material contact (the left boundary),  we have $\hat{u} = 0$ and $p = P_\mathrm{g}$. Thus, the flux is given by
\begin{equation}
\boldsymbol{F} = \frac{P_\mathrm{g}}{L}\begin{bmatrix}
0 \\ 1 \\ \dot{R}
\end{bmatrix}.
\end{equation}
For the boundary representing the shock front (the right boundary), the flux is given by
\begin{equation}
\boldsymbol{F} = \frac{1}{L}\begin{bmatrix}
-\rho_\infty \dot{R}_s \\ P_\infty \\ -\dot{R}_sE_\infty
\end{bmatrix},
\end{equation}
where $E_0 = (P_\infty + \gamma P_w)/(\gamma-1)$ is the total energy of the undisturbed fluid. Note that when the dynamic effect of the internal gas is considered, the Euler equation for $0<r<R$ should also be mapped to a scaled space and solved numerically. In such cases, its right boundary conditions can be connected to the conditions on the left boundary of the external fluid with a consistent flux.

conditions for the right boundary can be connected to the conditions for the left boundary of the external fluid with a consistent flux.
\subsection{Initial condition for the unified bubble theory}\label{sect:ic_for_bubble}
In this work, we start the simulation with the unified bubble theory at the moment denoted by $t_\mathrm{e}$ when $R_\mathrm{s}$ reaches $10R$. The initial condition at $t = t_\mathrm{e}$ is derived from the simulation of the initial phase prior to $t_\mathrm{e}$ using the equations above.
Let us denote $t = t_\mathrm{e}$ as the moment when the near-field simulation ends. Then, it is straightforward that the initial condition for the unified bubble theory is chosen as
\begin{equation}
\begin{cases}
P_0 = P_{\mathrm{g}}{(t_\mathrm{e})}, \\
R_0 = R{(t_\mathrm{e})},  \\
\dot{R}_0 = \dot{R}{(t_\mathrm{e})}.
\end{cases}
\end{equation}
However, the energy of the oscillating bubble will be excessive, and the maximum radius overestimated if the above is adopted directly. The key reason is that the energy of the shock wave which should have dissipated has been included. In this paper, we prevent this by calculating $\dot{R}_0$ with the kinetic energy of the fluid field excluding the shock wave region:

\begin{equation}
\dot{R}_0^2 = \frac{L}{R_0^3\rho_\infty}\int_0^a \rho u^2 r^2 d\zeta,
\end{equation}
where $a$ is a constant smaller than 1 to avoid the shock wave region in the integration. In this paper, $a$ is taken as 0.65.
\subsection{Shock wave far-field propagation}\label{sect-far-field}
Besides the bubble, the shock wave is also a major component of the physical process of an underwater explosion. The method to calculate near-field shock wave has been given in Appendix \ref{sect-near-field}. Nevertheless, 
when the far-field shock wave pressure is considered,  this method 
can be very expensive because of the time step limitation related to sound speed. Thus, we propose a new method based on the weakly compressible assumption for  quick estimation of far-field shock wave pressure. For far-field flow, 
we modeled the fluid with the Tait equation
\begin{equation}
p = (P_\infty+P_w) \left(\frac{\rho}{\rho_\infty}\right)^\gamma  - P_w.
\end{equation}
Thus, the sound speed $C$ satisfies
\begin{equation}
C^2 = \frac{\gamma}{\rho_\infty}(P_\infty+P_w )^{1/\gamma}(p+P_w)^{1-1/\gamma}.
\end{equation}
Expanding $C$ at $p = P_\infty$ and denoting $\delta p = p - P_\infty$,  we have
\begin{equation}\label{A22}
C =  C_\infty + K_c \delta p+ O(\delta p)^2,
\end{equation}
where $K_c = (\gamma-1)/(2\rho_\infty C_\infty)$.

The energy of the shock wave denoted by $E$ can be evaluated by $E = E_k + E_p \approx (\rho_\infty C_\infty^2)^{-1} (\delta p)^2$ in the far field. The shock wave energy is transported at the sound speed $C$. Thus, the conservation of the shock wave energy can be expressed as
\begin{equation}
\frac{\partial E}{\partial t} + \frac{\partial CE}{\partial r} = -\frac{2C}{r}E.
\end{equation}
Considering that the change of wavelength of the shock wave is slow in the far field, we map $r$ to $\zeta$ with the following relation
\begin{equation}
\zeta = \frac{r-R_\mathrm{e}}{R_\mathrm{s}-R_\mathrm{e}},
\end{equation}
where $R_\mathrm{e}$ is the left end 
of the computational domain and ${R_\mathrm{s}-R_\mathrm{e}}$ is the length of the computational domain which is a constant and denoted by $L_p$.
Approximating $u$ by $(\rho C)^{-1}\delta p$
, we have
\begin{equation}
\label{ae_at}
\frac{\hat{\partial}E}{\hat{\partial} t}  + \frac{1}{L_p}\frac{\partial (C-\dot{R}_s)E}{\partial \zeta} = -\frac{2 C}{r}E.
\end{equation}
Let's replace $E$ with $(\rho_\infty C_\infty^2)^{-1} p^2$, and $C$ with \ref{A22} neglecting $O(\delta p)^2$, 
then equation (\ref{ae_at}) can be expressed in a conservation-law form as
\begin{equation}
\frac{\hat{\partial}(\delta p)}{\hat{\partial} t} 
+ \frac{1}{L_p}(C_\infty-\dot{R}_s)\frac{\partial (\delta p)}{\partial \zeta} 
+ \frac34\frac{1}{L_p} K_c\frac{\partial (\delta p)^2}{\partial \zeta}
= -\frac{C}{r}\delta p.
\end{equation}
The above equation describes the evolution of a pressure wave in a spherical symmetric problem.
The second term on the left-hand side represents the linear transport of the wave energy. The third term is a nonlinear flux to reduce the slope of the pressure distribution curve after the shock front. The source term on the right-hand side incorporates the effect of the increase in the area of the shock front propagating outward. 
We can rewrite the above equation as
\begin{equation}\label{eq-shockwave-far}
\frac{\hat{\partial}(\delta p)}{\hat{\partial} t} + \frac{1}{L_p}\frac{\partial G}{\partial \zeta}
= S,
\end{equation}
where the flux $G = (C_\infty-\dot{R}_s)\delta p+ \frac34 K_c (\delta p)^2$ and the source $S = -C\delta p/r$. Then, the pressure history at $r$ can be expressed as
\begin{equation}
p(r,t) = \begin{cases}
P_\infty &\text{ when } r>R_\mathrm{s}, \\
P_\infty + \delta p &\text{ for else.}
\end{cases}
\end{equation}

The initial conditions of the far-field simulation are determined by the final state of the near-field simulation. 
In this work, we take the far-field as $L>10R$ where 
we assume that the shock wave pressure is low enough for the linear assumption to become valid. Therefore, we start calculation for the far-field using the approach in this section instead of the one based on the Euler equation introduced in Appendix \ref{sect-near-field}.  
The L2 projection is used to calculate the pressure solution in the new polynomial space from the solution of the Euler equation. 
The boundary conditions are given by the following boundary flux
\begin{equation}
\begin{cases}
\tilde{G} = G{(R_\mathrm{e})} \text{ for the left end, and } \\
\tilde{G} = G{(R_\mathrm{s})} \text{ for the right end.}
\end{cases}
\end{equation}

\subsection{Discontinuous Galerkin method}
The discontinuous Galerkin method \citep{DG_1d_system,DG_pp_limiter} is adopted to solve the near-field flow (the initial phase of bubble expansion) and the far-field shock wave propagation. The computational domain is discretized into $N_{dg}$ elements and the $\ell$th element is bounded by $\zeta_\ell$ and $\zeta_{\ell+1}$.

Firstly, we take the control equation (\ref{eq-shockwave-near}) of the near-field flow in Appendix \ref{sect-near-field} as an example. 
We approximate the solution by
\begin{equation}
\boldsymbol{U}_{(\zeta,t)} = \sum_{i=1}^{K+1} \beta_i({\zeta)} \boldsymbol{k}_{i(t)},
\end{equation}
where $\beta$ is a complete set of the polynomial space $P^K$.
For element $\ell$, the control equation (\ref{eq-shockwave-near}) of the near-field shock wave is then given in the Galerkin form
\begin{equation}
\int_{\zeta_\ell}^{\zeta_{\ell+1}} \beta_i\beta_jd\zeta \frac{\partial \boldsymbol{k}_j}{\partial t} = 
\int_{\zeta_\ell}^{\zeta_{\ell+1}} \left(\boldsymbol{F}\frac{\partial \beta_i}{\partial \zeta} + \boldsymbol{S}\beta_i \right)d\zeta - \left.\tilde{\boldsymbol{F}}\beta_i\right|_{\zeta=\zeta_\ell}^{\zeta=\zeta_{\ell+1}},
\end{equation}
where $\tilde{\boldsymbol{F}}$ is the flux at the boundary to be calculated according to specific boundary conditions.
The matrix form is given by
\begin{equation}
\dot{\boldsymbol{K}} = \boldsymbol{M}^{-1}\boldsymbol{B},
\end{equation}
where $M_{ij} = \int_{\zeta_\ell}^{\zeta_{\ell+1}} \beta_i \beta_j d\zeta$ and $\boldsymbol{B}_i = \int_{\zeta_\ell}^{\zeta_{\ell+1}} \left(\boldsymbol{F}\frac{\partial \beta_i}{\partial \zeta} + \boldsymbol{S}\beta_i\right)d\zeta - \left.\tilde{\boldsymbol{F}}\beta_i\right|_{\zeta=\zeta_\ell}^{\zeta=\zeta_{\ell+1}}$. The numerical flux $\tilde{\boldsymbol{F}}$ at $\zeta_\ell$ with $2\le\ell\le N_{dg}$  is calculated with the HLLC \citep{HLLC_Water} Riemann solver.

An explicit time marching scheme can be used to update $\dot{\boldsymbol{K}}$, e.g., the Runge-Kutta method. In this paper, 1 element with the $P^{15}$ space was used first until $L > 2R$. Then, 20 elements with the $P^3$ space were used until $L>10R$. If the air blast problem is considered, the Euler equation describing the explosion product should also be solved and the relay point between the near-field and far-field solvers should be later than the above for an underwater explosion problem because the compressibility of the air is much greater than that of the water. 

Subsequently, one may calculate the initial conditions for the unified bubble theory based on Appendix \ref{sect:ic_for_bubble} to analyze the bubble motion or start the far-field simulation based on Appendix \ref{sect-far-field} to predict the shock wave at a far distance. When solving the far-field shock wave propagation, equation (\ref{eq-shockwave-far}) is solved with the same approach. Only 1 element and $P^8$ space are used since the profile of the shock wave is smooth enough.
If the initial conditions of multi-bubble are going to be obtained at the same time and the interaction of their shock waves are considered, a full 3-Dimensional model should be used.

\subsection{Validation}


\begin{figure*}
	\centering\includegraphics[width=0.9\linewidth]{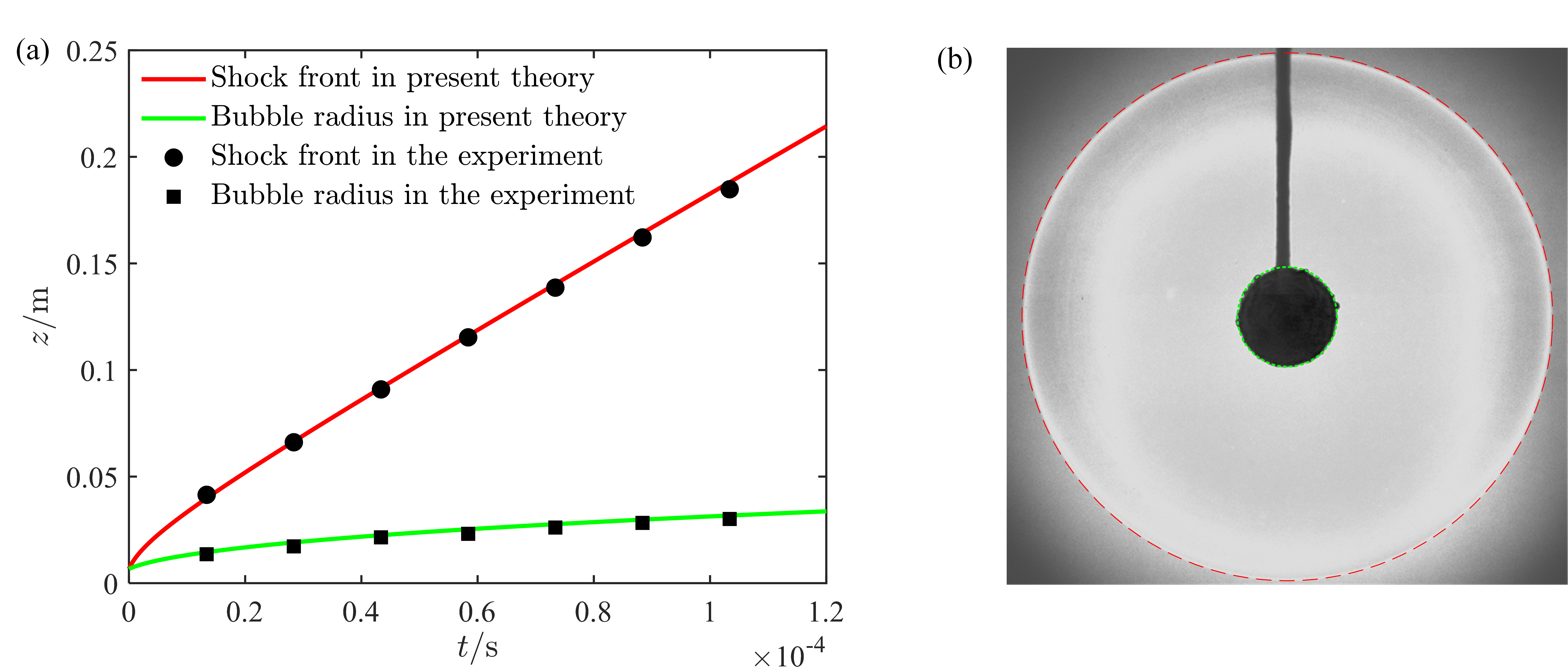}
	\caption{{Experimental and theoretical results at the initial phase of an underwater explosion with the charge equivalent to 2.0 g TNT. (a) Initial bubble expansion and shock wave front propagation. (b) An image of the underwater explosion captured at 0.106 ms from detonation. The theoretical results are projected on the image with the green and the red lines representing the bubble wall and the shock wave front, respectively.}}
	\label{shockfront}
\end{figure*}

\begin{figure*}
	\centering\includegraphics[width=1.0\linewidth]{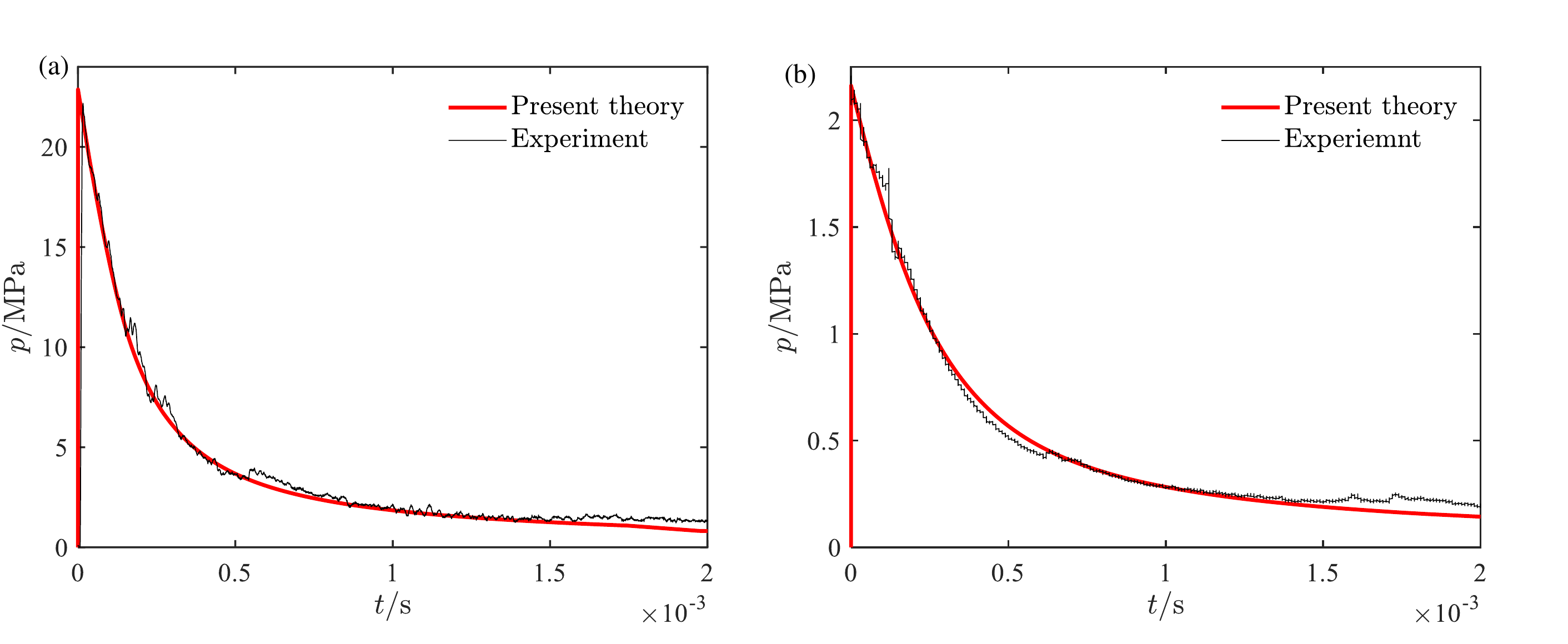}
	\caption{{Experimental and theoretical results of shock wave pressures for underwater explosions with the charges being equivalent to (a) 7.0 kg TNT measured at a distance of 4.0 m from the charge center and (b) 8.0 kg TNT measured at a distance of 30.0 m from the charge center.}}
	\label{shockpressure}
\end{figure*}

{Here, we validate the above methods 
	with experimental measurements. Three underwater explosion experiments were carried out in a free field with explosives equivalent to 2.0 g, 7.0 kg, and 80.0 kg TNT, respectively. We captured the shock wave front and the initial bubble expansion in the first experiment using the high-speed camera, to which we compared our solution. A good match was obtained as depicted in Fig.\ref{shockfront}. We also measured the detonation shock wave pressure histories in the second and the third experiments using a piezoelectric pressure sensor at distances of 4.0 m and 30.0 m, respectively. The comparison of the experimental and the simulated results using the near-field and far-field methods are illustrated in Fig. \ref{shockpressure} where a good agreement is achieved.}

\newpage
\bibliography{sample}

\end{document}